\documentclass[journal]{IEEEtran}

\usepackage{graphicx,cite,amsmath,amssymb,hhline,booktabs,stfloats,bm,amsthm,hyperref,times,subfigure}
\usepackage{algorithm,algorithmic,multicol,threeparttable,dcolumn,latexsym,rotating,textcomp,colortbl,enumerate}

\begin{document}

\title{Unified Offloading Decision Making and Resource Allocation in ME-RAN}
\author{
\bigskip
\medskip {\normalsize $\mbox{Kezhi Wang}^{}$, {\em Member, IEEE}, $\mbox{Pei-Qiu Huang}^{}$, {\em }$\mbox{Kun Yang}^{}$, {\em Senior Member, IEEE} }, $\mbox{Cunhua Pan}^{}$, {\em Member, IEEE}, $\mbox{Jiangzhou Wang}^{}$, {\em Fellow, IEEE}

\thanks{Copyright (c) IEEE. Personal use of this material is permitted. However, permission to use this material for any other purposes must be obtained from the IEEE by sending a request to pubs-permissions@ieee.org. }
\thanks{This work was supported in part by National Natural Science Foundation of China (Grant No. 61620106011 and 61572389); UK EPSRC NIRVANA project (Grant No. EP/L026031/1) and EU Horizon 2020 iCIRRUS project (Grant No. GA-644526).}
\thanks{Kezhi Wang is with Department of Computer and Information Sciences, Northumbria University, NE1 8ST, Newcastle upon Tyne, U.K, (e-mail: kezhi.wang@northumbria.ac.uk); Pei-Qiu Huang is with the School of Automation, Central South University, Changsha 410083, China, (e-mail: pqhuang@csu.edu.cn); Kun Yang is with School of Computer Sciences and Electrical Engineering, University of Essex, CO4 3SQ, Colchester, U.K and also with University of Electronic Science and Technology of China, Chengdu, China (e-mail: kunyang@essex.ac.uk); Cunhua Pan is with School of Electronic Engineering and Computer Science, Queen Mary University of London, London E1 4NS, U.K, (e-mails: C.Pan@qmul.ac.uk); Jiangzhou Wang is with School of Engineering and Digital Arts, University of Kent, Canterbury,
	Kent, CT2 7NZ, U.K, (e-mail: J.Z.Wang@kent.ac.uk).}
}
\markboth{IEEE Transactions on Vehicular Technology}%
{K Wang \MakeLowercase{\textit{et al.}}: Unified Offloading Decision Making and Resource Allocation in ME-RAN}

\maketitle

\vspace{-1cm}
\begin{abstract}
In order to support communication and computation cooperation, we propose ME-RAN architecture, which consists of mobile edge cloud (ME) as the computation provision platform and radio access network (RAN) as the communication interface. Cooperative offloading framework is proposed to achieve the following tasks: (1) to increase user equipment' (UE') computing capacity by triggering offloading action, especially for the UE which cannot complete the computation locally; (2) to reduce the energy consumption for all the UEs by considering limited computing and communication resources. Based on above objectives, we formulate the energy consumption minimization problem, which is shown to be a non-convex mixed-integer programming. Firstly, Decentralized Local Decision Algorithm (DLDA) is proposed for each UE to estimate the possible local resource consumption and decide if offloading is in its interest. This operation will reduce the overhead and signalling in the later stage. Then, Centralized decision and resource Allocation algoRithm (CAR) is proposed to conduct the decision making and resource allocation in ME-RAN. Moreover, two low complexity algorithms, i.e., UE with largest saved energy consumption accepted first (CAR-E) and UE with smallest required data rate accepted first (CAR-D) are proposed. Simulations show that the performance of the proposed algorithms is very close to the exhaustive search but with much less complexity.

\textbf{Intex Terms} - Communication and Computation Cooperation, Unified Offloading Decision Making, Resource Allocation, ME-RAN.
\end{abstract}

\section{Introduction}

Nowadays, user equipments (UEs) like smartphones and hand-held terminals are enjoying increasing popularity. However, due to limited resources in terms of battery, CPU, storage, etc, UEs are struggling in keeping up with the development of the resource intensive applications.

Mobile cloud computing (MCC) \cite{6195845,5445167,6923537} was proposed to make UEs with computing intensive tasks be able to offload computations to the cloud to increase UEs' experience and prolong their battery life.
Several cloud offloading platforms have been studied before, such as ThinkAir \cite{6195845}, which can migrate the applications from the mobile devices to the cloud. In \cite{6787113}, MCC has been applied to execute the offloaded computations and a game theoretic approach has been proposed to make the decision for each UE about where to execute the computation. However, the above mentioned MCC systems applied the normal cloud, such as Amazon elastic compute cloud (EC2) \cite{Amazon}, to execute the offloaded computations. In this case, UEs have to send their instructions, along with the data all the way via the Internet to the cloud. This is not beneficial to the UEs with high communication reliability and low latency requirement.

Mobile edge computing (MEC) \cite{MEC}, by moving a step further, proposes to set the cloud in the network edge. It can significantly reduce the latency of the task execution. Also, this technology is especially welcomed by the network operators, as it can make them go beyond from just the pipe providers, but also the cloud service operators. Furthermore, the operator has the potential to provide better cloud services to the UE than the normal cloud, as the mobile operator not only holds the computing information from the cloud, but also has the wireless channel status such that they can better jointly leverage both communication and computing resource.

Another cloud-based network infrastructure, i.e., cloud radio access network (C-RAN)
has also attracted operator's attention recently \cite{6998041, China, Pan}. C-RAN moves most of the network computation related tasks to central baseband unit (BBU) pool and distribute low complexity remote radio heads (RRHs) to the whole cell.
Due to the centralized management, signals from other UEs can be coordinated and are no longer considered as detrimental interference but useful signals.
Because of the centralized processing feature in C-RAN, it is of much interest to set edge cloud right next to RAN side, managed by the mobile operator.
In such a case, computing and communication resources may be monitored and processed together and bring not only good service to the UEs but also increase the profit for the mobile operators.

Earlier works on resource allocation and task scheduling in C-RAN with MEC, either consider there is only one UE conducting offloading, such as \cite{5445167}, or consider there is no interference between each offloaded tasks, such as \cite{8274943}. However, in wireless access channel, whether one UE decides to offload or not will induce interference to other UEs and affect other UEs' decisions, as the interference may deteriorate other UEs' signals. Some UEs may increase their transmission power to guarantee the high data rate and reliable transmissions. This action may in turn lead to the failure of the other UE's offloading packet transmission. Moreover, some other works, such as \cite{7105959, 7393804, 8353131}, assumed that the operator always has enough computing resource for all the offloaded UEs or for their communication requirement. However, computation resources are normally limited by the number of available physical machines. Therefore, admission control is normally necessary in managing the offloaded tasks and different level of priority may be imposed to the offloaded UEs. To the best of our knowledge, joint decision making and communication/computation resource allocation for multi-user offloading system considering interference has yet to be tackled, especially when communication and computation resource are limited.

In this paper, by applying C-RAN and MEC, mobile edge cloud-radio access network (ME-RAN) architecture is proposed. ME-RAN is composed of the mobile edge cloud (ME) and RAN. ME hosts both mobile clone (MC) and BBU, where MC and BBU are both implemented by cloud-based virtual machines. In ME-RAN, UEs with computation intensive task can offload it to the MC, whereas BBU is in charge of signal processing related tasks, such as receiving the computations from the UEs in the uplink and returning the results back to UEs. We aim to minimize the total energy consumption of all the UEs, by deciding the offloading set, the energy consumption for each UE (either offloading or conducting the tasks locally), the resource allocation and the receiving beamforming vectors in ME-RAN. The energy consumption minimization problem is formulated to be a mixed-integer non-convex programming, which is hard to solve in general. Exhaustive search is normally applied in this kind of problem but with prohibitive complexity.
In this paper, the whole offloading framework is established and low complexity algorithms are proposed with the main contributions summarized as follows:
\begin{itemize}
	\item  To reduce the signaling overhead and traffic between UE and ME-RAN, decentralized local decision algorithm (DLDA) is first proposed for
	each UE to estimate its possible local resource consumption and then decide if offloading is needed. Estimation model of energy consumption without knowing other UEs' decision and corresponding interference is provided. Only UE with offloading request will participate in the resource competition. This operation can be seen as the pre-screening of the offloading candidates.
	\item To tackle the obstacle that each UE itself does not have the global information when conducting offloading, Centralized decision and resource Allocation algoRithm (CAR) is proposed to be conducted by ME-RAN to make the decision on which UE can be allowed to offload and the corresponding resource allocation. Offloading priority is given to UEs which cannot complete the task locally. Uplink-downlink duality is employed to establish a link between offloading action from UE side and the available computing and communication resource from ME-RAN. The non-smooth indicator constraint is approximated as a non-convex function and the successive convex approximation (SCA) is applied to deal with this non-convexity. Also, auxiliary variables are applied to make the problem feasible to be solved in ME-RAN.
	\item Moreover, two low complexity algorithms, i.e., UE with largest saved energy consumption accepted first (CAR-E) and UE with smallest required data rate accepted first (CAR-D) are proposed to quickly conduct the decision making and resource allocation for each UE. The algorithms do not need any complex iteration.	
	\item Simulation results show that with the help of ME-RAN, most of the UEs which previously may not be able to execute the tasks locally now can not only complete the task in required time, but also enjoy high computation resource in edge cloud. Also, total energy consumption of all the UEs can be saved to a large extent compared to other traditional algorithms. Also, simulation shows that the performance of the proposed algorithm is very close to the exhaustive search but with much less complexity.
\end{itemize}

%\begin{comment}
The remainder of this paper is organized as follows. Section~\ref{Section:SystemModel} introduces the whole architecture design and system model of the proposed ME-RAN.
Section \uppercase\expandafter{\romannumeral3} analyses the problem and introduces the local pre-screening algorithm, i.e., DLDA. Section
\uppercase\expandafter{\romannumeral4} presents
the proposed centralized algorithms, i.e., CAR, followed by Section \uppercase\expandafter{\romannumeral5} with two low complexity algorithms, i.e., CAR-E and CAR-D.
Simulation results are presented in Section
\uppercase\expandafter{\romannumeral6}, whereas conclusion is made in Section \uppercase\expandafter{\romannumeral7}.

%\end{comment}
Notations: $\mathbb E (x)$ denotes the expectation of $x$, $\mathcal{CN}(0, \sigma^2 \mathbf{I} )$ denotes the complex Gaussian distribution with zero mean and covariance vector $\sigma^2 \mathbf{I}$, 's.t.' is short for 'subject to', the log function is the logarithm function with base 2, $|\cdot|$ denotes the size of the set, $|\cdot|_0$ is the indicator function defined in (\ref{ww21})
and $||\cdot||$ stands for either the Euclidean norm of a complex vector or the magnitude of a complex number, depending on the context.

\section{System Model}~\label{Section:SystemModel}
\subsection{Architecture}
\begin{figure}[t]
	\centering
	\includegraphics[width=3.3in]{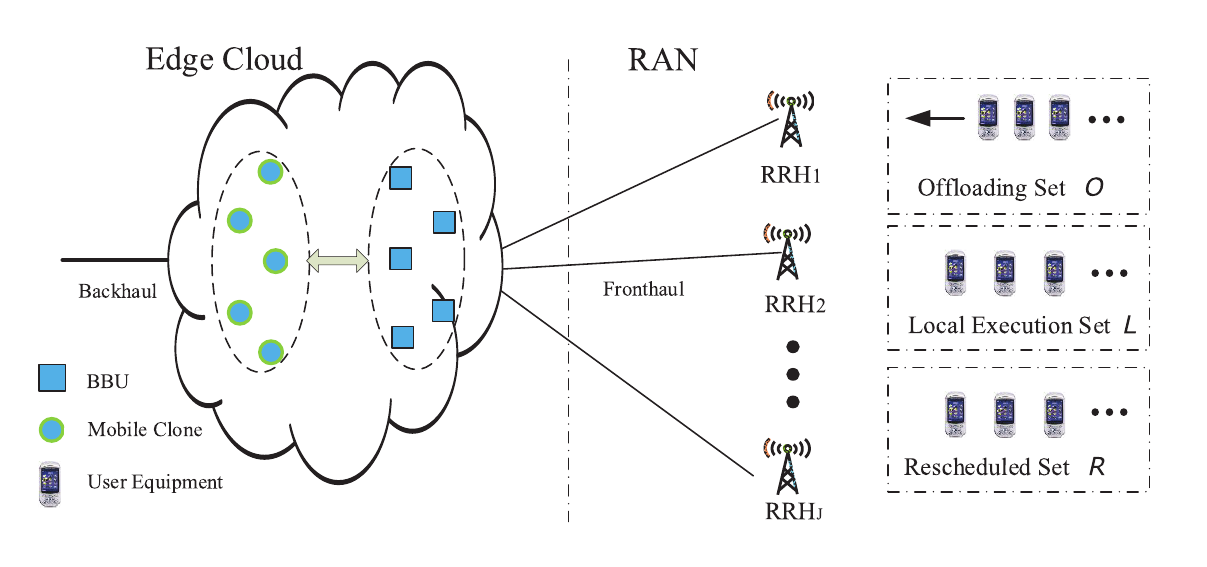}
	\caption{ME-RAN architecture.} \label{fig:Architecture}
\end{figure}

Assume that there is a ME-RAN network with $N$ UEs, each with one antenna, and $J$ RRHs, each of which has $K$ antennas connecting to the BBU pool through high-speed fiber fronthaul link, as shown in Fig.~\ref{fig:Architecture}. Denote the set of the UEs as $\mathcal{N}=\{1, 2,\cdots, N\}$ and the set of the RRHs as $\mathcal{J}=\{1, 2,\cdots, J\}$.

%Note that the analytical work can be extended to UEs with multiple antennas, where each multiple-antenna UE can be regarded as the combination of several virtual single-antenna UEs. Hence, all the derivations and algorithms developed in this paper can be generalized.

Similar to \cite{7393804}, it is assumed that each UE $i$ has the task $U_i$ to be accomplished as

\begin{equation}\label{wg5}
\begin{aligned}
U_i=(F_i, D_i, T_i), \forall i\in \mathcal{N},
\end{aligned}
\end{equation}
where $F_i$ (in cycles) describes the total number of the CPU cycles to be completed, $D_i$ (in bits) denotes the whole size of data required to be transmitted to ME-RAN if choosing to offload and $T_i$ (in seconds) is the delay constraint that this task has to be accomplished in order to satisfy the UE's quality of service (QoS) requirement.

In ME-RAN, each UE belongs to one of the following sets, according to its own status and current available computing and communication resources:
\begin{itemize}
	\item Offloading set $ \mathcal{O}$ is defined for UEs which have interests in offloading. $ \mathcal{O^H}$ represents the set for UEs which cannot complete the tasks locally and are assigned with high priority in offloading, while $ \mathcal{O^L}$ represents the set for the rest of UEs with offloading requests. Therefore, one has $ \mathcal{O}= \mathcal{O^H} \cup \mathcal{O^L}$.
	\item Local execution set $ \mathcal{L}$ is defined for UEs which decide to execute tasks locally.
	\item Rescheduled set $ \mathcal{R}$ is defined for UEs which neither complete the tasks locally due to lack of computing resource, nor offload due to lack of computing or communication resource.
\end{itemize}
Thus, one has $\mathcal{N}=\mathcal{L}\cup\mathcal{O} \cup \mathcal{R}$.

\subsection{Local Execution}~\label{Subsection:LocalExecution}
For the $i$-th UE which decides to conduct the task locally, i.e., $\forall i\in \mathcal{L}$, the execution time is
\begin{equation}\label{w3}
\begin{aligned}
T_i^L=\frac{F_i}{f_i^L}, \forall i\in \mathcal{L},
\end{aligned}
\end{equation}
where $f_i^L$ is the computation capability (i.e, CPU cycles per second) for the $i$-th UE.

Also, the computational power and the energy consumption can be given as \cite{6574874,Antti,6748976}
\begin{equation}\label{w4}
\begin{aligned}
p_i^L=\kappa_i^L (f_i^L)^{\nu^L_i}, \forall i\in \mathcal{L},
\end{aligned}
\end{equation}
where $\kappa_i^L > 0$ and $\nu^L_i \geq 2$ are the positive constants. According to the realistic measurements, $\kappa_i^L$ can be set to $\kappa_i^L=10^{-18}$ and $\nu^L_i$ can be set $\nu^L_i =3$. Then, the energy consumption is given by
\begin{equation}\label{w271}
\begin{aligned}
E_i^L =  p_i^{L} T_i^L, \forall i\in \mathcal{L}.
\end{aligned}
\end{equation}

By using latency requirement, one has
\begin{equation}\label{w5}
\begin{aligned}
C1: T_i^L \leq T_i, \forall i\in \mathcal{L}.
\end{aligned}
\end{equation}
Different UEs may have different computation capabilities and the constraints of $f_i^L$ is given by
\begin{equation}\label{w6}
\begin{aligned}
C2:  f_i^L \leq  f_{i,max}^L, \forall i\in \mathcal{L},
\end{aligned}
\end{equation}
where $f^L_{i,max}$ is the maximum computation capacity that the $i$-th UE can achieve and is finite.

\subsection{Task Offloading}~\label{Subsection:TaskOffloading}
For UEs who decide to offload the task, i.e., $\forall i\in \mathcal{O}$, the transmitted signal is written as
$
x_i=\sqrt{p_i^{Tr}} b_i, \forall i\in \mathcal{O}
$,
where $p_i^{Tr}$ denotes the transmission power of the $i$-th UE and $b_i$ denotes the transmitting data symbol with unity average power ${\mathbb E}(|b_i|^2)=1$.
Then, the received signal at the RRHs is given by
$
\mathbf{y}=\sum_{i\in \mathcal{O} } \mathbf{h}_{i}  \sqrt{p_i^{Tr}} b_i + \mathbf{z},
$
where $\mathbf{h}_i \in \mathcal{C}^{K \cdot J} $ denotes the channel state information (CSI) from $i$-th UE to all the RRHs, $\mathbf{z}$ denotes the additive white Gaussian noise (AWGN) vector and is assumed to be distributed as $\mathcal{CN}(0, \sigma^2 \mathbf{I} )$.
Then, the signal-to-interference-plus-noise ratio (SINR) can be expressed by
\begin{equation}\label{wk1}
\begin{aligned}
\text{SINR}^{UP}_i=\frac{p_i^{Tr} ||\mathbf{m}_i^H \mathbf{h}_i||^2}{\sum_{k \in \mathcal{O}, k\neq i} p_k^{Tr} || \mathbf{m}_i^H \mathbf{h}_k||^2 +\sigma^2 ||\mathbf{m}_i||^2}, \forall i\in \mathcal{O},
\end{aligned}
\end{equation}
where $\mathbf{m}_i \in \mathcal{C}^{K \cdot J}$ denotes the receive beamforming vector in RRHs for the $i$-th UE. Defining the maximum transmission power as $P_{i,max}$, we obtain
\begin{equation}\label{w18}
\begin{aligned}
C3: p_i^{Tr} \leq  P_{i,max},  \forall i\in \mathcal{O}.
\end{aligned}
\end{equation}
Thus, the achievable rate for the UE $i$ is given by
\begin{equation}\label{w11}
\begin{aligned}
r_i^{UP}= B \cdot \text{log}(1+\text{SINR}^{UP}_i), \forall i\in \mathcal{O},
\end{aligned}
\end{equation}
where $B$ is the wireless channel bandwidth.
If the $i$-th UE decides to offload the task to ME-RAN, the task data $D_i$ has to be transmitted to ME-RAN. From (\ref{w11}), the transmission time is given by
\begin{equation}\label{wk17}
\begin{aligned}
T_i^{Tr}=\frac{D_i}{r_i^{UP}}, \forall i\in \mathcal{O}.
\end{aligned}
\end{equation}
If the $i$-th UE decides to offload computation, the energy consumption is given by
\begin{equation}\label{w2171}
\begin{aligned}
E_i^{Tr} = p_i^{Tr} T_i^{Tr}, \forall i\in \mathcal{O}.
\end{aligned}
\end{equation}

\subsection{Mobile Edge Cloud (ME)}~\label{Subsection:ME}

It is assumed that ME hosts both mobile clone (MC) pool for service computation and BBU pool for communication computation.
\subsubsection{MC pool}~\label{Subsubsection:MCpool}
If the task is offloaded to mobile clone, the execution time in the $i$-th mobile clone can be expressed as
\begin{equation}\label{wh17}
\begin{aligned}
T_i^C=\frac{F_i}{f_i^C}, \forall i\in \mathcal{O},
\end{aligned}
\end{equation}
where $f_i^C$ is the computational capability of the $i$-th mobile clone. Then, the total time including data offloading and execution is given by
\begin{equation}\label{wff19}
\begin{aligned}
T_i^O=T_i^{Tr}+T_i^C, \forall i\in \mathcal{O}.
\end{aligned}
\end{equation}
As in \cite{6574874}, the time for sending data back to UE in the downlink is ignored.
Then, the following QoS constraints must hold
\begin{equation}\label{w18}
\begin{aligned}
C4: T_i^{O} \leq  T_i,  \forall i\in \mathcal{O}.
\end{aligned}
\end{equation}
%where QoS requirement $T_i$ is defined in (\ref{w1}).
Assuming that different mobile clones may have different computational capabilities and the constraint of the computation capacity of the $i$-th mobile clone is given by
\begin{equation}\label{w319}
\begin{aligned}
C5: f_i^C \leq  f^C_{i,max},  \forall i\in \mathcal{O},
\end{aligned}
\end{equation}
where $f^C_{i,max}$ is the maximum computation capacity that is allocated to the $i$-th mobile clone.

For the UE that is able to perform the computation locally, the condition for this UE to choose to offloading is that
\begin{equation}\label{w211171}
\begin{aligned}
C6: E_{i}^{Tr} \leq E_i^{L}, \forall i\in \mathcal{O^L}.
\end{aligned}
\end{equation}
In fact, $C6$ can imply that UE only considers offloading if its offloading energy consumption is smaller than its local execution.

Furthermore, the number of mobile clones is normally constrained by the number of virtual machines (or the number of CPU cores in the physical machines). Therefore, one has
\begin{equation}\label{ww219}
\begin{aligned}
{{{\left| {{\mathcal{O} }} \right|}}} \leq F^C,
\end{aligned}
\end{equation}
where $F^C$ is the maximal number of mobile clones which can be offered by the MC pool.
One can also rewrite (\ref{ww219}) as
\begin{equation}\label{w219}
\begin{aligned}
C7: \sum\limits_{i \in {\cal O}} {{{\left| {{{\left\| {{{\bf{m}}_i}} \right\|}^2}} \right|}_0}}   \leq F^C,
\end{aligned}
\end{equation}
where $\sum\nolimits_{i \in {\cal O}} {{{\left| {{{\left\| {{{\bf{m}}_i}} \right\|}^2}} \right|}_0}}$ stands for the number of offloading UEs and
\begin{equation}\label{ww21}
\begin{aligned}
{\left| {{{\left\| {{{\bf{m}}_i}} \right\|}^2}} \right|_0}=\left\{\begin{matrix}
&\;\;\;\;\;0,  \text{if} \ {{{\left\| {{{\bf{m}}_i}} \right\|}^2}} =0, \\& 1,  \text{otherwise}.
\end{matrix}\right.  % \; \forall i\in \mathcal{O}
\end{aligned}
\end{equation}
Thus, $C7$ implies that the maximum number of UEs that the MC can serve is $F^C$.

\subsubsection{BBU pool}~\label{Subsubsection:BBUpool}
In \cite{6477581}, the architecture of general processing processor (GPP) based BBU pool was presented and it showed that the required computational resource of BBU is influenced by the number of served UEs. Studies \cite{7105959,7864795} have shown that the computational capability in BBU is affected by the data rate of the serving UEs.
Therefore, one can assume the computational resource required in BBU pool is as
\begin{equation}\label{w21}
\begin{aligned}
{f^B} = \sum\limits_{i \in {\cal O}} {{{\left| {{{\left\| {{{\bf{m}}_i}} \right\|}^2}} \right|}}}_0 r_i^{UP} U,
\end{aligned}
\end{equation}
where $U$ (in cycle/bit) describes how much computing resource is required in the BBU to process one bit data. Without loss of generality, we assume $U=1$ cycle/bit in this paper. Then one can have the practical constraint as
\begin{equation}\label{w22}
\begin{aligned}
C8: \sum\limits_{i \in {\cal O}} {{{\left| {{{\left\| {{{\bf{m}}_i}} \right\|}^2}} \right|}_0}}  r_i^{UP}   \leq  F^{B},
\end{aligned}
\end{equation}
where $F^{B}$ (in cycles/second) is the maximum computational capacity in the BBU pool.

Note that, in the ME-RAN, service computing capacity $F^C$ and communication computing capacity $F^{B}$ can be allocated and adjusted according to the requirements.

%If more network resource is needed, $F^{B}$ can be allocated more than $F^C$. On the other hand, if there are more resource hungry task requests, $F^C$ can be allocated more than $F^{B}$. $F^C$ and $F^{B}$ are important system parameters in the practical system design.

\subsection{Problem Formulation}~\label{Subsection:ProblemFormulation}
Define binary variables $s_i \in \{ 0,1\}, \forall i\in \mathcal{N}$, where $s_i=1$ denotes that the UE chooses to offload, $s_i=0$ represents UE decides to compute the task locally. Then, one can formulate the energy minimization problem for all the UEs as
\begin{equation}\label{wqwwwe32}
\begin{aligned}
\mathcal{P}1: \;\;\; &\underset{ \mathbf{s}, \mathbf{f}, \mathbf{p}^{Tr}, \mathbf{m} }{\text{min}} \;\;\; \sum_{i\in \mathcal{N}} \left(s_i E_i^{Tr}+ (1-s_i) E_i^L\right)
\\\text{s.t. }\;
&C1: T_i^L \leq T_i, \;  i \in \mathcal{L},\\
&C2:  f_i^L \leq  f_{i,max}^L,  \; i \in \mathcal{L}, \\
&C3: p_i^{Tr} \leq  P_{i,max},  i\in \mathcal{O},\\
&C4:  T_i^{O} \leq  T_i, \;  i \in \mathcal{O}, \\
&C5: f_i^C \leq  f^C_{i,max}, \;   i \in \mathcal{O}, \\
&C6:  E_i^{Tr} \leq  E_i^L, \; i \in \mathcal{O^L},\\
&C7:  \sum\limits_{i \in {\cal O}} {{{\left| {{{\left\| {{{\bf{m}}_i}} \right\|}^2}} \right|}_0}}   \leq F^C,  \\
&C8:  \sum\limits_{i \in {\cal O}} {{{\left| {{{\left\| {{{\bf{m}}_i}} \right\|}^2}} \right|}_0}}  r_i^{UP} \leq F^B, \\
&C9:  s_i \in \{0,1\}, i \in \mathcal{N},   \\
\end{aligned}
\end{equation}
where $\mathbf{s}$, $\mathbf{f}$, $\mathbf{p}^{Tr}$ and $\mathbf{m}$ are the collection of all the decision variables, the allocated computing resource (including $f_i^{C}$ and $f_i^{L}$), the power consumption for all the UEs and
the receive beamforming vectors in ME-RAN, respectively. One can see that the sets $\mathcal{O}$, $\mathcal{L}$ and $\mathcal{O^L}$ are also variables, which will be decided during the process of our proposed algorithm next.
One can also see that $\mathcal{P}1$ is a non-convex mixed-integer programming, as the decision variable $\mathbf{s}$ is binary and $\mathbf{f}$, $\mathbf{p}^{Tr}$ and $\mathbf{m}$ are continuous.
%Normally, this kind of problem is NP-hard \cite{pochet2006production},
The exhaustive search may be applied to this problem, but with very high complexity.

\section{Distributed Local Decision Algorithm (DLDA)}~\label{Section:DLDA}

\subsection{Analysis to $\mathcal{P}1$}~\label{Subsection:P1}
From $\mathcal{P}1$, one can have the following observations:
\begin{itemize}
	\item Due to $C1 - C2$, not all the UEs are able to complete the tasks locally. For those UEs which cannot complete the tasks themselves, they have to seek MC for help.	
	Thus, offloading priority has to give to the UEs which can not conduct the tasks. In this case, $C6$ is no longer needed.
	\item Due to $C3 - C4$, not all the UEs are able to offload their computations, as the required transmission power may be larger than the maximal power capability of the UEs.
	\item Due to $C6$, not all the UEs in set $\mathcal{O^L}$ are willing to offload the tasks, as the required transmission energy consumption may be larger than their local executing energy consumption. Therefore, those UEs can be removed from the final offloading set. However, it is difficult for UE itself to know how much energy it needs to offload, as it is affected by the decisions of other UEs and the interference caused by them.
	\item Due to the limitation of the available resources in ME-RAN, i.e., $C5$, $C7$ and $C8$, not all the offloading requests from the UEs can be accepted by ME-RAN. Thus, access control has to be imposed to decide the feasible offloading set.
	\item  One can also notice that $C7$ may determine the maximum number of UEs allowed to offload, whereas $C5$ and $C8$ may determine which UEs can be allowed to offload, based on their required offloading resource.
\end{itemize}

Based on the above observation, we propose the following offloading protocol:
\begin{itemize}
	\item  Step 1: Each UE conducts local decision (i.e., DLDA, introduced in Section III. B) to decide if offloading is to its interest, based on local information, such as channel state information and processing capacity, etc.
	Only those UEs that meet the offloading criteria send the requests to the ME-RAN for resource competition.
	\item  Step 2: Based on the offloading requests receiving from the UEs, ME-RAN will conduct centralized decision and resource allocation algorithm (CAR, introduced in Section IV or CAR-E and CAR-D, introduced in Section V). CAR (or CAR-E and CAR-D) will decide which UEs can be allowed to offload, how much power each UE can apply and the resource allocation. Then ME-RAN sends the above instructions to each UE.
	\item  Step 3: Each UE follows the instructions received from ME-RAN and proceed to offload, such as applying the required offloading power.
\end{itemize}
In above protocol, Step 1 can decrease the overhead and traffic between UE and ME-RAN in wireless channel, as UE which does not see the offloading benefit will not send the offloading request. This can also reduce the complexity of central decision in Step 2, as the variable space of central decision is reduced.
Note that after each UE receives the instructions from ME-RAN in step 2 (i.e., the offloading power, etc), all the UEs are required to follow the instructions and apply the corresponding resource allocation, e.g., adjusting its offloading power. Similar assumption has been made in references, such as \cite{6884852}.

One may reformulate $\mathcal{P}1$ as
\begin{equation}\label{wqwww32}
\begin{aligned}
&
 \underset{ \mathbf{s}, \mathbf{f}, \mathbf{p}^{Tr}, \mathbf{m} }{\text{min}} \;\;\; \sum_{i\in \mathcal{N}} \left( s_i (  E_i^{Tr}-E_i^L )+E_i^L \right)
\\\text{s.t. }\;
&\text{Constraints of}\;  \mathcal{P}1.
\end{aligned}
\end{equation}
The above problem (\ref{wqwww32}) may be further reformulated as
\begin{equation}\label{wq2wewq2}
\begin{aligned}
&\underset{ \mathbf{s}, \mathbf{f}, \mathbf{p}^{Tr}, \mathbf{m} }{\text{min}} \;\;\; \sum_{i\in \mathcal{O}}  E_i^{Tr}  - M \cdot  \sum_{i\in \mathcal{N}}s_i
\\\text{s.t. }\;
&\text{Constraints of}\;  \mathcal{P}1,
\end{aligned}
\end{equation}
where $M$ is a very large value and $\sum_{i\in \mathcal{N}}s_i$ is the number of offloading UEs.

\textbf{Proof}:
By using $C6$, one can have $E_i^{Tr}-E_i^L \leq 0$. Also $E_i^L$ is the constant. Therefore, increasing $s_i$ will further reduce the objective value of (\ref{wqwww32}). Furthermore, as $E_i^{Tr}$ is a positive value, reducing the sum of $E_i^{Tr}$ will also reduce the objective value of (\ref{wqwww32}). By using the large value of $M$ in (\ref{wq2wewq2}), we can first increase the number of offloaded UEs as many as possible. At the same time, we reduce the total energy consumption for all the offloaded UEs. Therefore, one can see that minimizing the objective of (\ref{wqwww32}) can be possibly done by minimizing objective of (\ref{wq2wewq2}). Although this transformation is not optimal, it is especially useful to the practical system, as normally the UEs would like to offload the tasks to the cloud to save their local resource if possible (or if there is available resource).   \qed

\subsection{Distributed Local Decision Algorithm (DLDA)}
Before we show the DLDA in Algorithm 1, some propositions are presented first.

\textbf{Proposition 1}: If UE conducts the task itself, the optimal CPU frequency is given by $f_i^{L^*}=\frac{F_i}{T_i} $ and the local
energy consumption is given as
$ E_i^{L^* }  = \kappa_i^L \frac{F_i^{\nu^L_i}}{T_i^{\nu^L_i-1}}$. If $f_i^{L^*}> f_{i,max}^L $, this UE can not complete the task locally, and has to offload the task to cloud.

\textbf{Proof}: For each UE conducting the task itself, the minimization of energy consumption can be written as
\begin{equation}\label{w136}
\begin{aligned}
\mathcal{P}1.1 \;\;\; &\underset{f_i^L}{\text{min}} \;\;\;   E_i^L,\; \forall i \in \mathcal{N}
\\\text{s.t. }\;  &C1, C2.
\end{aligned}
\end{equation}
For above problem, as the delay constraint for the task is $T_i$, one can easily obtain the optimal clock frequency as
$f_i^{L^*}=\frac{F_i}{T_i}$, the optimal local executing power as $p_i^{L^* }  = \kappa_i^L \frac{F_i^{\nu^L_i}}{T_i^{\nu^L_i}}$ and the local
energy consumption as
$ E_i^{L^* }  = \kappa_i^L \frac{F_i^{\nu^L_i}}{T_i^{\nu^L_i-1}}$.
However, the above solution is only feasible if
$ f_i^{L^*} \leq f_{i,max}^L$
and there is no solution if
$
f_i^{L^*}  >   f_{i,max}^L,
$
which means the minimum clock frequency required for executing the task locally is larger than the maximum clock frequency available at this UE. \qed

\textbf{Proposition 2}: If UE decides to offload, the minimal  transmission power is
\begin{equation}\label{w50}
\begin{aligned}
p_{i,min}^{Tr}=\frac{ \left( 2^{\frac{R_{i,min}}{  B}}-1 \right) \sigma^2  }{  || \mathbf{h}_i||^2}.
\end{aligned}
\end{equation}

\textbf{Proof}:
The minimal transmission power is determined by the minimum achievable rate.
By using $C4$ and $C5$, one can get the minimum achievable rate as
\begin{equation}\label{w47}
\begin{aligned}
C10:  r_i^{UP} \geq R_{i,min},
\end{aligned}
\end{equation}
where
\begin{equation}\label{w48}
\begin{aligned}
R_{i,min}=\frac{D_i}{T_i-\frac{F_i}{f^C_{i,max}}}.
\end{aligned}
\end{equation}
Then, from (\ref{wk1}) and (\ref{w11}), one can get the transmission power as
\begin{equation}\label{w49}
\begin{aligned}
p_{i,min}^{Tr'}=\frac{ \left( 2^{\frac{R_{i,min}}{  B }} -1 \right)  \iota  }{  ||\mathbf{m}_i^H \mathbf{h}_i||^2},
\end{aligned}
\end{equation}
where $\iota =\sum_{k \in \mathcal{O}, k\neq i} p_k^{Tr} || \mathbf{m}_i^H \mathbf{h}_k||^2 +\sigma^2 ||\mathbf{m}_i||^2$.

The minimal transmission power can be obtained by assuming there is only one UE conducting offloading, i.e., no interference from other UEs.
By applying channel-matched decoding vector, one can get the minimal transmitting power as (\ref{w50}). \qed

Note that from (\ref{w50}),  the minimum transmit power can be calculated based only on local information, since the required information is available at each UE. Then, based on above analysis, we propose the local decision DLDA conducted in each UE to initially decide $\mathcal{O^L}$, $\mathcal{O^H}$, $\mathcal{R}$ and $\mathcal{L}$,  without any global information.

\begin{algorithm}[t]
	\caption{Decentralized Local Decision Algorithm (DLDA)}
	\footnotesize
	\begin{algorithmic}[1]
		\STATE Each UE $\forall i \in \mathcal{N}$ obtains $f_i^{L^*}$ and $E_i^{L^* }$ by solving problem $\mathcal{P}1.1$;
		\IF{$f_i^{L^*}  >   f_{i,max}^L$}
		\STATE Add this UE into $\mathcal{O^H}$;
		\ELSE
		\STATE Add this UE into $\mathcal{\mathcal{N}\backslash O^H}$;
		\ENDIF
		\STATE Each UE $\forall i \in \mathcal{O^H}$ obtains $p_{i,min}^{Tr}$ from (\ref{w50});
		\IF{$p_{i,min}^{Tr}> P_{i,max} $}
		\STATE Move this UE from $\mathcal{O^H}$ to $\mathcal{R}$ and update $\mathcal{O^H}$;
		\ENDIF
		\STATE Each UE $ \forall i \in \mathcal{\mathcal{N}\backslash O^H} \backslash \mathcal{R}$ obtains $p_{i,min}^{Tr}$ from (\ref{w50});
		\IF{$p_{i,min}^{Tr}> \text{min}\left(P_{i,max}, P^\Delta  _{i,max} \right)$ }
		\STATE Add this UE into $\mathcal{L}$;
		\ELSE
		\STATE Add this UE into $\mathcal{O^L}$;
		\ENDIF
		\STATE \bf{Output}  $\mathcal{O^L}$, $\mathcal{O^H}$, $\mathcal{L}$, $\mathcal{R}$ to central decision.
	\end{algorithmic}
\end{algorithm}

DLDA is summarized in Algorithm $1$, where each UE first checks if it can complete task locally, by solving $\mathcal{P}1.1$. The UEs that cannot complete the tasks will be assigned high priority (adding them to $\mathcal{O^H}$). Otherwise, they can be added to $\mathcal{\mathcal{N}\backslash O^H}$.
For the UEs in the set $\mathcal{O^H}$, if their minimal transmission power obtained from (\ref{w50}) is larger than the maximal power, i.e., $p_{i,min}^{Tr}> P_{i,max}$, then they will be moved from $\mathcal{O^H}$ to $\mathcal{R}$. For the UEs in the set $ \forall i\in \mathcal{\mathcal{N}\backslash O^H} \backslash \mathcal{R}$, if the UEs are not able to offload the tasks, i.e., $p_{i,min}^{Tr}> P_{i,max}$,  then the UEs will be added into set $\mathcal{L}$ as well.
For the rest of UEs in set $\forall i\in \mathcal{\mathcal{N}\backslash O^H} \backslash \mathcal{R}$, they can be added into offloading set $\mathcal{O^L}$ as the offloading candidates, which will be updated in central decision next based on available resource in MC-RAN.

For the UE finally accepted by MC-RAN to offload task, its transmission power needs to meet the following constraint (\ref{w52}), otherwise this UE may not have interest in offloading.

\begin{equation}\label{w52}
\begin{aligned}
p_i^{Tr}\leq
\text{min}\left(P_{i,max}, P^\Delta  _{i,max} \right),
\end{aligned}
\end{equation}
where $ P^\Delta  _{i,max}= \frac{E_i^{L^*}}{T_i-\frac{F_i}{f^C_{i,max}}} $ and $E_i^{L^*}$ is given by {\textbf{Proposition 1}.

\textbf{Proposition 3}: For each UE choosing to offload, minimizing its offloading energy consumption is equivalent to minimizing its transmitting power consumption.
	
\textbf{Proof}:
The transmission energy minimization of each UE can be formulated as
\begin{equation}\label{w1136}
\begin{aligned}
\mathcal{P}1.2: \;\;\; &\underset{p_i^{Tr}}{\text{min}} \;\;\;   E_i^{Tr},\; \forall i \in \mathcal{N}
\\\text{s.t. }\;  &p_{i,min}^{Tr}\leq p_i^{Tr}\leq
\text{min}\left(P_{i,max}, P^\Delta  _{i,max} \right).
\end{aligned}
\end{equation}

For $\mathcal{P}1.2$, the objective function can be written as
\begin{equation}\label{w44}
\begin{aligned}
E^{Tr}_i=\frac{p_i^{Tr} D_i}{B\text{log}(1+\frac{p_i^{Tr} ||\mathbf{m}_i^H \mathbf{h}_i||^2}{\sum_{k \in \mathcal{O}, k\neq i} p_k^{Tr} || \mathbf{m}_i^H \mathbf{h}_k||^2 +\sigma^2 ||\mathbf{m}_i||^2})}.
\end{aligned}
\end{equation}
	
By taking the derivative of $E^{Tr}_i$ with respect of $p_i^{Tr}$, one can get
\begin{equation}\label{w45}
\begin{aligned}
&\frac{\partial E^{Tr}_i}{\partial p_i^{Tr} }=\frac{\lambda(p_i^{Tr}) }{B (\kappa p_i^{Tr}+\iota ) \log ^2\left(\frac{\kappa p_i^{Tr}}{\iota }+1\right)},
\end{aligned}
\end{equation}
where $\lambda(p_i^{Tr}) = D_i \left((\kappa p_i^{Tr}+\iota ) \log \left(\frac{\kappa p_i^{Tr}}{\iota }+1\right)-\kappa p_i^{Tr}\right)$, $\kappa = ||\mathbf{m}_i^H \mathbf{h}_i||^2$, and $\iota$ is given in \eqref{w49}.
	
Since the denominator in~\eqref{w45} is greater than 0, we only need to take the numerator, i.e., $\lambda(p_i^{Tr})$, into account. By taking the derivative of $\lambda(p_i^{Tr})$ with respect of $p_i^{Tr}$, one can get
\begin{equation}\label{w46}
\begin{aligned}
&\frac{\partial \lambda(p_i^{Tr})}{\partial p_i^{Tr} }=D_i\kappa \log\left(\frac{\kappa p_i^{Tr}}{\iota }+1\right) > 0.
\end{aligned}
\end{equation}
Therefore, one can see that $\lambda(p_i^{Tr})$ is an increasing function with respect to $p_i^{Tr}$. As $p_{i,min}^{Tr} > 0$, one has $\lambda(p_{i}^{Tr}) \geq \lambda(p_{i,min}^{Tr}) > \lambda(0)=0$. Therefore, one can have $\frac{\partial E^{Tr}_i}{\partial p_i^{Tr} } > 0$, which implies $E^{Tr}_i$ increases with the increase of $p_i^{Tr}$. Therefore, the minimal transmission energy can be obtained if the minimal transmission power is applied. \qed

\textbf{Proposition 4}: For each UE choosing to offload, the minimal transmitting energy is given by 
\begin{equation}\label{w501}
\begin{aligned}
E_{i,min}^{Tr*}=p_{i,min}^{Tr}\left(T_i-\frac{F_i}{f_i^{C*}}\right),
\end{aligned}
\end{equation}
where the optimal computational resource is given as $f_i^{C*} = f_{i,max}^C$.

\textbf{Proof}:
From \eqref{w50} and \eqref{w48}, one can see that if $f_i^C = f_{i,max}^C$, the minimal transmission power can be obtained. Then, from \textbf{Proposition 3}, $f_i^{C*} = f_{i,max}^C$ must also hold for obtaining UE's 
minimal offloading energy. This is intuitive as that all the UEs normally try to use all the possible computing resource from cloud to save their energy consumption. Therefore, the optimal computational resource allocated to each offloaded UE can be obtained as $f_i^{C*} = f_{i,max}^C$. \qed

In the next step, UEs in sets $\mathcal{O^H}$ and $\mathcal{O^L}$ will send offloading requests to ME-RAN for communication and computation resource competition. UEs in set $\mathcal{O^H}$ will be given high priority. As the resource may be limited, access control is conducted. If being declined by ME-RAN, UEs in set $\mathcal{O^L}$ will be moved to set $\mathcal{L}$ and conduct tasks locally, whereas UEs in set $\mathcal{O^H}$ will be moved to set $\mathcal{R}$ and postpone the task execution to the next time slot.

\section{Centralized Decision and Resource Allocation Algorithm (CAR)}
After receiving the offloading requests from UEs in sets $\mathcal{O} =\mathcal{O^H} \cup \mathcal{O^L}$ obtained in last section, CAR is conducted in ME-RAN to decide the final offloading set and resource allocation.

From (\ref{wq2wewq2}), one can see that we first need to maximize the number of offloading tasks. Then, minimize the total power for all offloading UEs. Thus, access control may be imposed to the offloaded UEs.
Three cases can be considered:
\begin{itemize}
	\item Case I: The communication and computation resource in ME-RAN is large enough to accommodate all the offloaded UEs (i.e., $ {\left|\cal O  \right|}\leq F^C$ and $\sum\limits_{k \in {\cal O}}  R_{k,min} \leq F^B  $). Thus, no access control is needed.
	\item Case II: The communication and computation resource in ME-RAN is not enough to accommodate the offloaded UEs even with high priority (i.e., for the UEs in set $\mathcal{O^H}$). Thus, access control is imposed to the UEs in set $\mathcal{O^H}$. No offloading is allowed for the UEs in set $\mathcal{O^L}$.	
	\item Case III: The communication and computation resource in ME-RAN is enough to accommodate the offloaded UEs with high priority (i.e.,  $ {\left|\cal O^H  \right|}\leq F^C$ and $\sum\limits_{k \in {\cal O^H}}  R_{k,min} \leq F^B  $). Thus, access control is only imposed to the UEs in set $\mathcal{O^L}$.
\end{itemize}
Thus, one can summarize the proposed CAR as Algorithm 2.

\begin{algorithm}[t]
	\caption{Centralized Decision and Resource Allocation Algorithm (CAR)}
	\label{alg5}
	\begin{algorithmic}[1]
		\IF { $ {\left|\cal O  \right|}\leq F^C$ and $\sum\limits_{k \in {\cal O}}  R_{k,min} \leq F^B  $ }
		\STATE \textbf{run} Case I in Algorithm 3;
		\ELSIF{ $ {\left|\cal O^H  \right|}\leq F^C$ and $\sum\limits_{k \in {\cal O^H}}  R_{k,min} \leq F^B  $ }
		\STATE \textbf{run} Case III in Algorithm 5;
		\ELSE
		\STATE \textbf{run} Case II in Algorithm 4.		
		\ENDIF
	\end{algorithmic}
\end{algorithm}
Next we will introduce how we deal with above Case I, Case II and Case III in Algorithm 3, Algorithm 4 and Algorithm 5, respectively.

\subsection{Case I}
If the resource is large enough, then no access control is needed and ME-RAN will accept all the requests from UEs. Thus, we only have to deal with the resource allocation problem.
In this case, $C7$ as well as $C8$ can be removed and
(\ref{wq2wewq2}) can be rewritten as
\begin{equation}\label{wrww32}
\begin{aligned}
&\underset{\mathbf{p}^{Tr}, \mathbf{m} }{\text{min}} \;\;\;  \sum_{i\in \mathcal{O}} E_i^{Tr}
\\\text{s.t. }\;
&C6, C10.
\end{aligned}
\end{equation}

According to {\bf{Proposition 3}}, for each UE, reducing the power consumption
is equivalent to minimizing its energy consumption. Therefore, we assume all the UEs would like to reduce its own power consumption when we consider to minimize the energy consumption of all the UEs as a whole (i.e., (\ref{wrww32})).
In fact, saving each UE's power assumption can also reduce the interference to other UEs, thereby saving energy consumption of other UEs. As a result, energy consumption minimization in \eqref{wrww32} can be transformed to power minimization as
\begin{equation}\label{wrww33}
\begin{aligned}
&\underset{\mathbf{p}^{Tr}, \mathbf{m} }{\text{min}} \;\;\;  \sum_{i\in \mathcal{O}} p_i^{Tr}
\\\text{s.t. }\;
&C6, C10.
\end{aligned}
\end{equation}

Similar to \cite{1453766}, we consider (\ref{wrww2}) first, and then check if the individual energy consumption constraint $C6$ can be met (i.e., if offloading can save UE's energy consumption).
\begin{equation}\label{wrww2}
\begin{aligned}
&\underset{\mathbf{p}^{Tr}, \mathbf{m} }{\text{min}} \;\;\;  \sum_{i\in \mathcal{O}} p_i^{Tr}
\\\text{s.t. }\;
&C10.
\end{aligned}
\end{equation}
For (\ref{wrww2}), one can have its dual problem as
\begin{equation}\label{ww32}
\begin{aligned}
&\underset{ \mathbf{v} }{\text{min}} \;\;\;  \sum_{i\in \mathcal{O}} \mathbf{v}_i^H \mathbf{v}_i
\\\text{s.t. }\;
& C11: r_i^{VD} \geq R_{i,min}, \;\forall i \in \mathcal{O},
\end{aligned}
\end{equation}
where $\mathbf{v}_i \in \mathcal{C}^{K \cdot J} $ is the virtual downlink transmission beamforming vector from all the RRHs to $i$-th UE, $\mathbf{v}$ is a collection of all the $\mathbf{v}_i$,
$r_i^{VD}$ is the virtual downlink transmission data rate defined as
$
r_i^{VD}= B \cdot \text{log}(1+\text{SINR}^{VD}_i)
$
and
$
\text{SINR}^{VD}_i=\frac{|| \mathbf{h}_i^H\mathbf{v}_i||^2}{\sum_{k\in \mathcal{O},\; k\neq i} ||  \mathbf{h}_i^H\mathbf{v}_k||^2 +\sigma^2}$ \cite{6884852}.
Assume $\mathbf{m^*}$, $\mathbf{p^{Tr*}}$ and $\mathbf{v^*}$ as the optimal solutions to problems (\ref{wrww2}) and (\ref{ww32}), respectively. Then similar to \cite{1453766}, $\mathbf{v^*}$ and $\mathbf{m^*}$ can be set to be identical and moreover, one can have
$\sum_{i\in \mathcal{O}}  p_i^{Tr}  =\sum_{i\in \mathcal{O}}  \mathbf{v}_i^H \mathbf{v}_i$ in above problems. Also, similar to \cite{6884852}, for any given feasible solution to problem (\ref{ww32}), one can always find a corresponding feasible solution to problem (\ref{wrww2}), and vice versa. Therefore, problems (\ref{wrww2}) and (\ref{ww32}) can take the same optimal value with the same set of beamforming vectors, i.e., $\mathbf{v^*}$ and $\mathbf{m^*}$ can be set to be identical.

In problem (\ref{ww32}), $C11$ can be transformed to the second-order cone (SOC) constraint in the virtual downlink as \cite{1561584}
\begin{equation}\label{www32}
\begin{aligned}
\sqrt{1-\frac{1}{2^{ \frac{R_{i,min} }{B }}}}\sqrt{\sum_{k\in \mathcal{O}}|| \mathbf{h}_i^H\mathbf{v}_k||^2 +\sigma^2}
\leq \text{Re}\left(\mathbf{h}_i^H\mathbf{v}_i\right).
\end{aligned}
\end{equation}
Therefore, (\ref{ww32}) becomes
\begin{equation}\label{w2w32}
\begin{aligned}
&\underset{ \mathbf{v} }{\text{min}} \;\;\;  \sum_{i\in \mathcal{O}} \mathbf{v}_i^H \mathbf{v}_i
\\\text{s.t. }\;
& (\ref{www32}), \;\forall i \in \mathcal{O}.
\end{aligned}
\end{equation}
One can see that (\ref{w2w32}) is a convex problem which can be solved efficiently, i.e., using interior point method.
Then similar to \cite{6884852}, by setting $\mathbf{m}=\mathbf{v}$ and using fixed-point method in (\ref{wrww2}), $\mathbf{p}^{Tr}$ can be obtained.

Then, we define a new set $\mathcal{B}_1$ that includes UEs in set $\mathcal{O^L}$ whose energy consumption are larger than their local energy consumption. Then one can have $\mathcal{B}_1=\{  i| E_i^{Tr^*} >  E_i^{L^*},  i \in \mathcal{O^L}\}$.
Define a set of the normalized violation factor for each user's energy consumption in $\mathcal{B}_1$ as $\{\eta_i=\frac{E_i^{Tr^*}-E_i^{L^*}}{E_i^{L^*}}, i \in \mathcal{O^L}\}$. If $\mathcal{B}_1$ is not empty, some UEs in $\mathcal{B}_1$ may be moved from $\mathcal{O^L}$ to $\mathcal{L}$ and execute locally.
Our idea is to first remove the $i^*$-th UE with the biggest normalized violation factor, i.e., $i^*=\text{argmax} (\eta _i, i\in \mathcal{B}_1 )$ from set $\mathcal{O^L}$ and then redo problem (\ref{w2w32}) again until $\mathcal{B}_1=\varnothing $. Therefore, we can summarize the process to solve Case I in Algorithm 3.

\begin{algorithm}
	\caption{Case I in CAR.}
	\label{alg2}
	\begin{algorithmic}[1]
		\STATE Obtain $\mathbf{p}^{Tr}$ from (\ref{wrww2}) and (\ref{ww32}) and obtain $\mathcal{B}_1=\{  i| E_i^{Tr^*}>  E_i^{L^*},  i \in \mathcal{O^L}\}$;
		\IF {$\mathcal{B}_1\neq \varnothing $}
		\STATE Order $\eta_i=\frac{E_i^{Tr^*}-E_i^{L^*}}{E_i^{L^*}}$, $i\in \mathcal{B}_1$ and find the largest $i^*=\text{argmax} (\eta _i, i\in \mathcal{B}_1 )$ and remove $i^* $-th UE from $\mathcal{O^L}$ and add it into $\mathcal{L}$, \textbf{go to} step 1
		\ENDIF
		%		\STATE Update $\mathcal{L}$ and $\mathcal{O^L}$
	\end{algorithmic}
\end{algorithm}

\subsection{Case II}

If the resource is not large enough to accept the UEs with high offloading priority, access control is conducted to UEs in set $\mathcal{O^H}$. No UEs in set $\mathcal{O^L}$ are allowed to offload. In this case, $C6$ can be removed.
(\ref{wq2wewq2}) can be rewritten as

\begin{equation}\label{w32}
\begin{aligned}
&\underset{\mathbf{p}^{Tr}, \mathbf{m} }{\text{min}} \;\;\;  \sum_{i\in \mathcal{O^H}} E_i^{Tr}
\\\text{s.t. }\;
&C7, C8, C10.
\end{aligned}
\end{equation}

Similar with above, the problem can be rewritten as the following power minimization.

\begin{equation}\label{w33}
	\begin{aligned}
		&\underset{\mathbf{p}^{Tr}, \mathbf{m} }{\text{min}} \;\;\;  \sum_{i\in \mathcal{O^H}} p_i^{Tr}
		\\\text{s.t. }\;
		&C7, C8, C10.
	\end{aligned}
\end{equation}
Further, the dual problem can be written as

\begin{equation}\label{wr58}
\begin{aligned}
&\underset{\mathbf{v}}{\text{min}} \;\;\;  \sum_{i\in \mathcal{O^H}} \mathbf{v}_i^H \mathbf{v}_i
\\& \text{s.t.}: \; C11, C12: \sum\limits_{i \in {\cal O^H}} {{{\left| {{{\left\| {{{\bf{v}}_i}} \right\|}^2}} \right|}_0}}  \leq F^C,
\\& C13: \sum\limits_{i \in {\cal O^H}} {{{\left| {{{\left\| {{{\bf{v}}_i}} \right\|}^2}} \right|}_0}}  R_{i,min}\leq F^B.
\end{aligned}
\end{equation}
Again, similar to \cite{6884852}, we can see that for any given feasible solution to problem (\ref{w33}), one can always find a corresponding feasible solution to problem (\ref{wr58}), and vice versa. If we assume $\mathbf{m^*}$, $\mathbf{p^{Tr*}}$ and $\mathbf{v^*}$ as the optimal solutions to problems (\ref{w33}) and (\ref{wr58}), respectively, one can set $\mathbf{v^*}$ and $\mathbf{m^*}$ to be identical. Also, one can have
$\sum_{i\in \mathcal{O^H}}  p_i^{Tr}  =\sum_{i\in \mathcal{O^H}}  \mathbf{v}_i^H \mathbf{v}_i$ in above problems.

Then, we will focus on how to solve (\ref{wr58}). Two obstacles still avoid us to directly solve the problem because: 1) the feasibility of the problem is still unknown and 2) the non-smooth indicator functions in the constraints are hard to tackle. Next, we will show how to deal with the above two hurdles.

Inspired by \cite{4570234}, one can use nonnegative auxiliary variables in (\ref{wr58}) to deal with the feasibility problem, which can then be transformed to
\begin{equation}\label{w11166}
\begin{aligned}
&\underset{\mathbf{v}, \mathbf{y}}{\text{min}} \;\;\;  \sum_{i\in \mathcal{O^H}} \mathbf{v}_i^H \mathbf{v}_i +M  \sum_{i\in \mathcal{O^H}} y_i
\\& \text{s.t.}:\; C12, C13,
C14: \sqrt{1-\frac{1}{2^{ \frac{R_{i,min} }{B }}}}\sqrt{\sum_{k\in \mathcal{O^H}}|| \mathbf{h}_i^H\mathbf{v}_k||^2 +\sigma^2} \\&
\leq \text{Re}\left(\mathbf{h}_i^H\mathbf{v}_i \right) + y_i, \forall i \in \mathcal{O^H},
\end{aligned}
\end{equation}
where $ \{y_i,\; i\in \mathcal{O^H} \}$ are the nonnegative auxiliary variables and
$\mathbf{y}$ is a collection of $ \{y_i,\; i\in \mathcal{O^H} \}$. One can see that there always exists large enough variables $ \{y_i,\; i\in \mathcal{O^H} \}$ to satisfy all the constraints in above problem. By solving (\ref{w11166}), we can obtain the value of $ \{y_i,\; i\in \mathcal{O^H} \}$. The number of zero entries in $\{y_i,\; i\in \mathcal{O^H} \}$ in (\ref{w11166}) corresponds to the number of accepted UEs in set $\mathcal{O^H}$. Similarly, one can also obtain the set of the accepted UEs by checking
$\{  i| r_i^{VD}\geq  R_{i,min}, i \in  \mathcal{O^H} \}$.

Note that non-smooth indicator function $C12$ and $C13$ still makes (\ref{w11166}) intractable.
Fortunately, they can be approximated by applying the following fractional function, i.e.,
\begin{equation}\label{w62}
\begin{aligned}
f_\theta (x) = \frac{x}{x+\theta},
\end{aligned}
\end{equation}
where $\theta$ is a small positive value. One can see that a very small $x$ results in $f_\theta (x) \approx 0$, whereas a large $x$ leads to $f_\theta (x) \approx 1$.
Then $C12$ and $C13$ can be approximated as $C15$ and $C16$, respectively.
\begin{equation}\label{w63}
\begin{aligned}
C15: \sum_{i\in \mathcal{O^H}} f_\theta \left({\left\| {{{\bf{v}}_i}} \right\|^2}\right)  \leq F^C,
\end{aligned}
\end{equation}
\begin{equation}\label{w64}
\begin{aligned}
C16: \sum_{i\in \mathcal{O^H}} f_\theta \left({\left\| {{{\bf{v}}_i}} \right\|^2}\right)  R_{i,min}\leq F^B.
\end{aligned}
\end{equation}
In practice, we set $\theta=10^{-3}$, and if $\left\| {{{\bf{v}}_i}} \right\|^2< 10^{-3}$, one can set $\left\| {{{\bf{v}}_i}} \right\|^2=0$ to make above transformation feasible.

Then, by using $C15$ and $C16$, problem (\ref{w11166}) can be transformed into the following problem
\begin{equation}\label{w1616}
\begin{aligned}
&\underset{\mathbf{v}, \mathbf{y}}{\text{min}} \;\;\;  \sum_{i\in \mathcal{O^H}} \mathbf{v}_i^H \mathbf{v}_i +  M  \sum_{i\in \mathcal{O^H}} y_i
\\& \text{s.t.}:\; C14, C15, C16.
\end{aligned}
\end{equation}
Problem (\ref{w1616}) is more tractable than (\ref{w11166}), as both the objective function and constraints in (\ref{w1616}) are continuous and differentiable. Although Problem (\ref{w1616}) is still nonconvex due to the concavity of $f_\theta (\cdot)$ in $C15$ and $C16$, it is a well-known difference of convex (d.c.) program, which can be solved effectively by using the SCA method \cite{dinh2010local}. This approach was proposed to approximate the concave function as Taylor expansion with first order. Therefore, by using the concavity of $f_\theta (x)$, one can have
\begin{equation}\label{w65}
f_\theta (||\mathbf{v}_i||^2) \leq f_\theta (||\mathbf{v}_i{(t)}||^2)+ \alpha_i(t) ( ||\mathbf{v}_i||^2 -||\mathbf{v}_i{(t)}||^2),
\end{equation}
where $\mathbf{v}_i{(t)}$ is the solution of $i$-th UE in the $t$-th iteration, $\alpha_i(t)= f'_\theta (||\mathbf{v}_i{(t)}||^2)$ and $f'_\theta (x)$ is the first-order derivative of $x$. By replacing $f_\theta (\cdot)$ in
(\ref{w1616}) with the right hand side of (\ref{w65}), we can solve the following optimization in the $(t+1)^{th}$ iteration as
\begin{equation}\label{w1617}
	\begin{aligned}
	&\underset{\mathbf{v}, \mathbf{y}}{\text{min}} \;\;\;  \sum_{i\in \mathcal{O^H}} \mathbf{v}_i^H \mathbf{v}_i +  M  \sum_{i\in \mathcal{O^H}} y_i
	\\& \text{s.t.}:\; C14, C17: \sum_{i\in \mathcal{O^H}} \alpha_i(t) ||\mathbf{v}_i||^2 \leq F^C- \\& \sum_{i\in \mathcal{O^H}}\left(
	f_\theta (||\mathbf{v}_i{(t)}||^2)- \alpha_i(t)  ||\mathbf{v}_i{(t)}||^2 \right)
	\\& C18:  \sum_{i\in \mathcal{O^H}} R_{i,min} \alpha_i(t)  ||\mathbf{v}_i||^2     \leq F^B - \\& \sum_{i\in \mathcal{O^H}} \left(
	f_\theta (||\mathbf{v}_i{(t)}||^2)- \alpha_i(t) ||\mathbf{v}_i{(t)}||^2  \right).
	\end{aligned}
\end{equation}
One can see that (\ref{w1617}) is a convex problem, which can be solved by interior point method efficiently. The UE with the largest gap to its required data rate, i.e., $y_i$ in $C14$ are most likely to be forced to further reduce its virtual downlink transmission power to zero and encouraged
to drop out of $\mathcal{O^H}$ eventually. However, UE with smallest gap to its target data rate, such as $y_i=0$ in $C14$ will keep its virtual downlink transmission power non-zero and thus one can have $\left|||\mathbf{v}_i||^2\right|_0=1$ to indicate UE is accepted by ME-RAN. We summarize the process to solve Case II as Algorithm 4, where $\mathbf{v}(t)$ and $\mathbf{\alpha}(t)$ are the collection of $\mathbf{v}_i(t)$ and $\mathbf{\alpha}_i(t)$, $i \in \mathcal{O^H}$ respectively, in the $t$-th iteration.
\begin{algorithm}
	\caption{Case II in CAR.}
	\label{alg3}
	\begin{algorithmic}[1]
		\STATE \textbf{Initialize} $t=1$, $\mathbf{v}(0)$ and
		$\mathbf{\alpha}(0)$
		\WHILE{Convergence or pre-defined iterations reached}
		\STATE Solve $(\ref{w1617})$ to get $\mathbf{v}(t)$ with $\mathbf{v}(t-1)$, $\mathbf{\alpha}(t-1)$;  \\
		\STATE Update $\mathbf{\alpha}(t)$
		with $\mathbf{v}(t)$;\\
		\ENDWHILE
		\STATE  \textbf{Update} $\mathcal{O^H}=\{  i| r_i^{UP}\geq  R_{i,min}, i \in  \mathcal{O^H} \}$;
		\STATE \textbf{Update} $\mathcal{R}$ by adding $\{  i| r_i^{UP} <  R_{i,min}, i \in  \mathcal{O^H} \} $ into $\mathcal{R}$.
		%		\STATE Update $\mathcal{L}$ and $\mathcal{O^L}$
	\end{algorithmic}
\end{algorithm}

\subsection{Case III}
If the resource is not enough to accept all the offloaded UEs but is able to accommodate the UEs with high priority (i.e., UE $i \in  \mathcal{O^H}$), then access control is only imposed on the UEs in set $\mathcal{O^L}$.
Thus, after guaranteeing the resource in $\mathcal{O^H}$, the remaining resource in ME-RAN is given by
\begin{equation}\label{w67}
\begin{aligned}
C19: \sum_{i\in \mathcal{O^L}} f_\theta \left(||\mathbf{v}_i||^2\right)  \leq F^{C}-  \sum\limits_{k \in {\cal O^H}} {\left| {{{\left\| {{{\bf{m}}_k}} \right\|}^2}} \right|}_0,
\end{aligned}
\end{equation}
\begin{equation}\label{w68}
\begin{aligned}
C20: \sum_{i\in \mathcal{O^L}} f_\theta \left(||\mathbf{v}_i||^2\right)  R_{i,min}\leq F^B - \sum\limits_{k \in {\cal O^H}} {\left| {{{\left\| {{{\bf{m}}_k}} \right\|}^2}} \right|}_0 R_{k,min}.
\end{aligned}
\end{equation}
Then, (\ref{wq2wewq2}) can be rewritten as
\begin{equation}\label{w69}
\begin{aligned}
&\underset{\mathbf{v}, \mathbf{z}}{\text{min}} \;\;\;  \sum_{k\in \mathcal{O}} \mathbf{v}_k^H \mathbf{v}_k+M  \sum_{j\in \mathcal{O^L}} y_j
\\& \text{s.t.}:\; C19, C20, C21: \sqrt{1-\frac{1}{2^{ \frac{R_{m,min} }{B }}}}\sqrt{\sum_{k \in\mathcal{O}}|| \mathbf{h}_m^H\mathbf{v}_k||^2 +\sigma^2}
\\& \leq \text{Re}\left(\mathbf{h}_m^H\mathbf{v}_m \right) , \forall m\in \mathcal{O^H},
\\&
C22:\sqrt{1-\frac{1}{2^{ \frac{R_{j,min} }{B }}}}\sqrt{\sum_{k\in \mathcal{O}}|| \mathbf{h}_j^H\mathbf{v}_k||^2 +\sigma^2}
\\& \leq \text{Re}\left(\mathbf{h}_j^H\mathbf{v}_j \right) + y_j, j\in \mathcal{O^L},
\end{aligned}
\end{equation}
where $ \{y_i,\; i\in \mathcal{O^L} \}$ is a set of nonnegative auxiliary variables to ensure the feasibility of above problem, $\mathbf{y}$ is a collection of $ \{y_i,\; i\in \mathcal{O^L} \}$,
$C21$ is applied to guarantee the offloading date rate from $\mathcal{O^H}$ whereas
$C22$ is the relaxed constraint to guarantee the offloading date rate from $\mathcal{O^L}$. One can see that (\ref{w69}) is a convex problem which can be solved efficiently.
Similar with Case II, define a new set $\mathcal{B}_2$ that includes UEs in set $\mathcal{O^L}$ whose energy consumption are larger than their local energy consumption as $\mathcal{B}_2=\{  i| E_i^{Tr^*}> E_i^{L^*},  i \in \mathcal{O^L}\}$. Also, define a set of the normalized violation factor for each user's energy consumption as $\{\eta_i=\frac{E_i^{Tr^*}-E_i^{L^*}}{E_i^{L^*}}, i \in \mathcal{B}_2\}$. If $\mathcal{B}_2 \neq \varnothing$, one can
move $i^*$-th UE with $i^*=\text{argmax} (\eta _i, i\in \mathcal{B}_2 )$ from set $\mathcal{O^L}$ to set $\mathcal{L}$ and then redo problem (\ref{w69}) again until $\mathcal{B}_2=\varnothing $. One can summarize the process to solve Case III as Algorithm 5.

\begin{algorithm}
	\caption{Case III in CAR.}
	\label{alg4}
	\begin{algorithmic}[1]
		\STATE \textbf{Initialize} $t=1$, $\mathbf{v}(0)$ and
		$\mathbf{\alpha}(0)$
		\WHILE{Convergence or pre-defined iterations reached}
		\STATE Solve $(\ref{w69})$ to get $\mathbf{v}(t)$ with $\mathbf{v}(t-1)$, $\mathbf{\alpha}(t-1)$;  \\
		\STATE Update $\mathbf{\alpha}(t)$
		with $\mathbf{v}(t)$;\\
		\ENDWHILE
		\STATE Obtain $\mathbf{p}^{Tr}$ and $\mathcal{B}_2=\{  i| E_i^{Tr^*}>  E_i^{L^*},  i \in \mathcal{O^L}\}$;
		\IF {$\mathcal{B}_2\neq \varnothing $}
		\STATE Order $\eta_i=\frac{E_i^{Tr^*}-E_i^{L^*}}{E_i^{L^*}}$, $i\in \mathcal{B}_2$ and find the largest $i^*=\text{argmax} (\eta _i, i\in \mathcal{B}_2)$ and remove $i^* $-th UE from $\mathcal{O^L}$ and add it into $\mathcal{L}$, \textbf{go to} step 1;
		\ENDIF		
		\STATE \textbf{Update} $\mathcal{L}$ by adding $\{  i| r_i^{UP} <  R_{i,min}, i \in  \mathcal{O^L} \} $ into $\mathcal{L}$.
		%		\STATE Update $\mathcal{L}$ and $\mathcal{O^L}$
	\end{algorithmic}
\end{algorithm}

\section{Fast Decision and Resource Allocation Algorithm}

One can see that above Case II and Case III in CAR include iterations (i.e., while loop using interior point method), which may increase the complexity of the algorithm. In this section, we provide two fast CAR algorithms with low complexities, i.e., UEs with largest saved energy consumption accepted first (CAR-E) and UEs with smallest required data rate accepted first (CAR-D). Note that in this section, we only consider how to reduce the complexity for Case II and Case III in CAR in the last section, while the algorithm for Case I will be the same as before.

\subsection{UE with largest saved energy consumption accepted first (CAR-E)}
In this subsection, CAR-E is introduced, where UEs with largest saved energy consumption are accepted first.
The idea behind this algorithm is that we first assume there are enough resources (i.e., no constraints $C7$ and $C8$) and obtain the allocated power for each UE. Then, the UE with the maximum saved energy consumption will be accepted first until either $C7$ or $C8$ is violated. Therefore, (\ref{w1617}) can be rewritten as (\ref{w166}), if without considering the resource constraints $C7$ and $C8$.
\begin{equation}\label{w166}
\begin{aligned}
&\underset{\mathbf{v}, \mathbf{y}}{\text{min}} \;\;\;  \sum_{i\in \mathcal{O^H}} \mathbf{v}_i^H \mathbf{v}_i +  M  \sum_{i\in \mathcal{O^H}} y_i
\\& \text{s.t.}:\; C14.
\end{aligned}
\end{equation}
Then we first accept UE $i^*=\text{argmax} (E_i^{L^*}-E^{Tr^*}_i, i\in \mathcal{O^H} )$, until $C7$ or $C8$ is violated (i.e., the resource has all used up). UE which can not be accepted from $\mathcal{O^H}$ will be moved to $\mathcal{R}$ and execute in the next time slot.
Therefore, one can have the process for solving Case II by using CAR-E as in Algorithm 6.
\begin{algorithm}
	\caption{Case II in CAR-E.}
	\label{alg6}
	\begin{algorithmic}[1]
		\STATE Obtain $\mathbf{p}^{Tr}$ from (\ref{w166}) and (\ref{wrww2});
		\STATE  Initialize $\mathcal{O'^H} = \mathcal{O^H}$ and  $\mathcal{O^H}= \varnothing$;
		\WHILE {$C7$ and $C8$ are both met}
		\STATE Move UE $i^*=\text{argmax} (E_i^{L^*}-E^{Tr^*}_i, i\in \mathcal{O'^H })$ from set $\mathcal{O'^H}$ to $\mathcal{O^H}$;
		\ENDWHILE
		\STATE \textbf{Update} $\mathcal{R}$ by adding $\mathcal{O'^H}$ into $\mathcal{R}$.
		%		\STATE Update $\mathcal{L}$ and $\mathcal{O^L}$
	\end{algorithmic}
\end{algorithm}

Similarly, for Case III, (\ref{w69}) can be rewritten as (\ref{w6w29}), if without considering the resource constraints $C7$ and $C8$.
\begin{equation}\label{w6w29}
\begin{aligned}
&\underset{\mathbf{v}, \mathbf{z}}{\text{min}} \;\;\;  \sum_{k\in \mathcal{O}} \mathbf{v}_k^H \mathbf{v}_k+M  \sum_{j\in \mathcal{O^L}} y_j
\\& \text{s.t.}: C21, C22.
\end{aligned}
\end{equation}

Define the set for UEs whose energy consumption are smaller than their local energy consumption as $\mathcal{D}_1=\{  i| E_i^{Tr^*}\leq  E_i^{L^*},  i \in \mathcal{O^L}\}$ and also define a set of the normalized saved energy consumption as $\{\epsilon_i= \frac{ {E_i^{L^*}-E_i^{Tr^*}}}{E_i^{L^*}},  i \in \mathcal{D}_1\}$.} Then, our idea is to first accept the $i^*$-th UE with the biggest normalized saving power, i.e., UE $i^*=\text{argmax} (\epsilon _i, i\in \mathcal{D}_1 )$, until $C7$ or $C8$ is violated. UE which can not be accepted from $\mathcal{O^L}$ will be moved to $\mathcal{L}$ and execute the task locally.
Then, we can solve Case III by using CAR-E as in Algorithm 7.
\begin{algorithm}
	\caption{Case III in CAR-E.}
	\label{alg7}
	\begin{algorithmic}[1]
		\STATE Obtain $\mathbf{p}^{Tr}$ from (\ref{w6w29}) and (\ref{wrww2});
		\STATE Obtain $\mathcal{D}_1=\{  i| E_i^{Tr^*}\leq  E_i^{L^*},  i \in \mathcal{O^L}\}$ and $\{\epsilon_i= \frac{ {E_i^{L^*}-E_i^{Tr^*}}}{E_i^{L^*}},  i \in \mathcal{D}_1\}$;
		\STATE  Initialize $\mathcal{O'^L}= \mathcal{O^L}$ and $\mathcal{O^L}= \varnothing$;
		\WHILE {$C7$ and $C8$ are both met}
		\STATE Move UE $i^*=\text{argmax} (\epsilon _i, i\in \mathcal{D}_1 )$ from set $ \mathcal{D}_1$ to $\mathcal{O^L}$;
		\ENDWHILE
		\STATE \textbf{Update} $\mathcal{L}$ by adding $\mathcal{O'^L} \backslash  \mathcal{O^L}$ into $\mathcal{L}$.
	\end{algorithmic}
\end{algorithm}

\subsection{UE with smallest required data rate accepted first (CAR-D)}

In this subsection, CAR-D is proposed, where UE with smallest required data rate is accepted first. Similarly with above, we first assume that there are enough resources and then UEs with the smallest required data rate will be accepted sequentially until either $C7$ or $C8$ is violated. One can have the process to deal with Case II in CAR-D as Algorithm 8, where in step 4, one sees that we accept the UE with smallest required data rate first
\begin{algorithm}
	\caption{Case II in CAR-D.}
	\label{alg8}
	\begin{algorithmic}[1]
		\STATE Obtain $\mathbf{p}^{Tr}$ from (\ref{w166}) and (\ref{wrww2});
		\STATE  Initialize $\mathcal{O'^H} = \mathcal{O^H}$ and  $\mathcal{O^H}= \varnothing$;
		\WHILE {$C7$ and $C8$ are both met}
		\STATE Move UE $i^*=\text{argmin} (R_{i,min}, i\in \mathcal{O}'^H )$ from set $\mathcal{O'^H}$ to $\mathcal{O^H}$;
		\ENDWHILE
		\STATE \textbf{Update} $\mathcal{R}$ by adding ${\mathcal{O'^H}}$ into $\mathcal{R}$.
		%		\STATE Update $\mathcal{L}$ and $\mathcal{O^L}$
	\end{algorithmic}
\end{algorithm}

For Case III in CAR-D, similarly with before, we first define the set for UEs whose energy consumption is larger than its local energy consumption, i.e., $\mathcal{D}_2=\{i | {E_i^{L^*} \leq E_i^{Tr^*}} ,  i \in \mathcal{O^L}\}$ and then accept the UE in $\mathcal{D}_2$ with smallest required data rate first, until either $C7$ or $C8$ is violated. The whole process is shown in Algorithm 9.

\begin{algorithm}
	\caption{Case III in CAR-D.}
	\label{alg9}
	\begin{algorithmic}[1]
		\STATE Obtain $\mathbf{p}^{Tr}$ from (\ref{w6w29}) and (\ref{wrww2});
		\STATE Obtain $\mathcal{D}_2=\{  i| E_i^{Tr^*}\leq  E_i^{L^*},  i \in \mathcal{O^L}\}$
		\STATE  Initialize $\mathcal{O'^L} = \mathcal{O^L}$ and  $\mathcal{O^L}= \varnothing$;
		\WHILE {$C7$ and $C8$ are both met}
		\STATE Move UE $i^*=\text{argmin} (R_{i,min}, i\in \mathcal{D}_2 )$ from set $\mathcal{D}_2$ to $\mathcal{O^L}$;
		\ENDWHILE
		\STATE \textbf{Update} $\mathcal{L}$ by adding $\mathcal{O'^L} \backslash  \mathcal{O^L}$ into $\mathcal{L}$.	
	\end{algorithmic}
\end{algorithm}

\section{Simulation}
In this section, simulation are presented to show the effectiveness of the proposed algorithm.
Matlab with CVX tool \cite{cvx} has been applied.
\subsection{Setup }
\begin{figure}[htbp]
	\centering
	\includegraphics[width=2.5in]{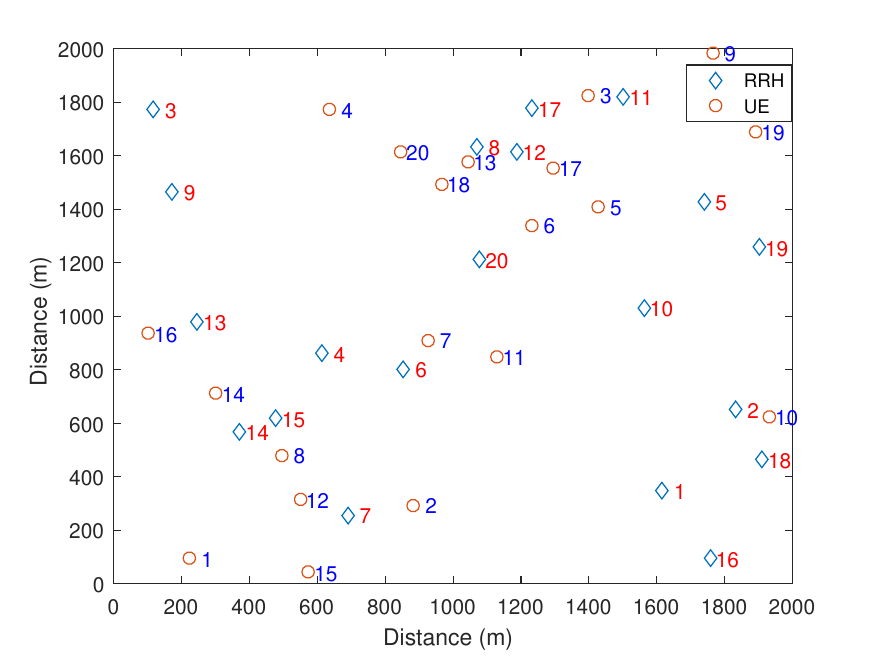}
	\caption{The simulation
		environment with $N=20$ UEs, each equipped with $K=2$ antennas and with $L=20$ RRHs, each equipped one antenna, assumed to be randomly distributed in a square area of coordinates [0, 2000] $\times$ [0, 2000] meters.} \label{mapfinal}
\end{figure}

The simulation
scenario is shown in Fig. \ref{mapfinal}, where
there are $N=20$ UEs, each with one antenna and $L=20$ RRHs, each equipped with $K=2$ antennas.
All the RRHs and UEs are assumed to be randomly distributed in a square area of coordinates [0, 2000] $\times$ [0, 2000] meters.
The path and penetration loss are assumed as
$p (d)=148.1+37.6\text{log}10(d)$,
where $d$ (km) is the propagation distance. It is assumed that the small scale fading is independent circularly symmetric Gaussian process distributed as $\mathcal{CN}(0, 1)$. The noise power spectral density is assumed to be $-75\; \text{dBm/Hz}$ and
the system bandwidth $B$ is set to 10 MHz, the maximum transmission power for each UE is set to 1W.
Moreover, the computation resource in each UEs $f^L_{i,min}$ is $10^6$ cycles/s, while the
maximum computation capacity for each mobile clone $f^C_{i,min}$ is $10^8$ cycles/s. The time slot $T_i$ is set to 1 s for all the UEs.
Unless noted otherwise, each UE has the computing task to be completed, with the transmission data ${D_i, \;i \in \mathcal{N}}$ and has CPU cycles required ${F_i, \;i \in \mathcal{N}}$ randomly selected as shown in Table I.
\begin{table*}% ???¨¨?????¨¨¡¤?????????¨¨??????????paper?????????
	\caption{\label{tab1}  Transmission data ${D_i, i \in \mathcal{N}}\; (\times 10^6 \;\text{bits})  $ and CPU cycles required ${F_i, \;i \in \mathcal{N}} \; (\times 10^6 \text{cycles} )$ for each UE.
	}
	\begin{center}
		{\begin{tabular}{|c|c|c|c|c|c|c|c|c|c|c|}% ?????????????¡¤????¨¦?????r:??????????????????¨¦?????c:?????????????¡À????
				\hline
				$D_1-D_{10}$    & 0.08  & 0.65 & 0.4& 0.15& 0.15& 0.4& 0.6& 0.7& 0.25& 0.6  \\
				\hline
				$D_{10}-D_{20}$ & 0.15& 0.69& 0.55& 0.56& 0.15& 0.65& 0.28& 0.19& 0.14& 0.25 \\
				\hline
				\hline
				$F_1-F_{10}$    & 0.2& 1 & 0.96&1.1& 0.8& 1.1& 0.8&  1.21& 1.4& 1.3  \\
				\hline
				$F_{10}-F_{20}$ & 1.1& 0.95& 0.9&    0.8&   1.08&  0.9&  0.75&    0.88&   0.95&   0.93  \\
				\hline
		\end{tabular}}
	\end{center}
\end{table*}
Next, we examine our proposed algorithm (i.e., \textbf{CAR}) and the low complexity algorithms (i.e., \textbf{CAR-E} and \textbf{CAR-D}), with the following solutions for comparison.

\begin{itemize}
	\item \textbf{Local Execution (Local)}: All the UEs execute the tasks locally.
	\item \textbf{Exhaustive Search (ES)}: We check all the possibilities, with the objective of minimizing the sum energy consumption for all the UEs.
\end{itemize}
Note that for UEs neither conduct the tasks locally, nor offload to the cloud, i.e., in set $\mathcal{R}$, we still calculate their local energy consumption by using \textbf{Proposition 1} and assuming $f_i^{L^*}= f_{i,max}^L $ in the simulation, which is only for the convenience of the comparison.

\subsection{Results and Insights}

\begin{figure}[htbp]
	\centering
	\includegraphics[width=2.5in]{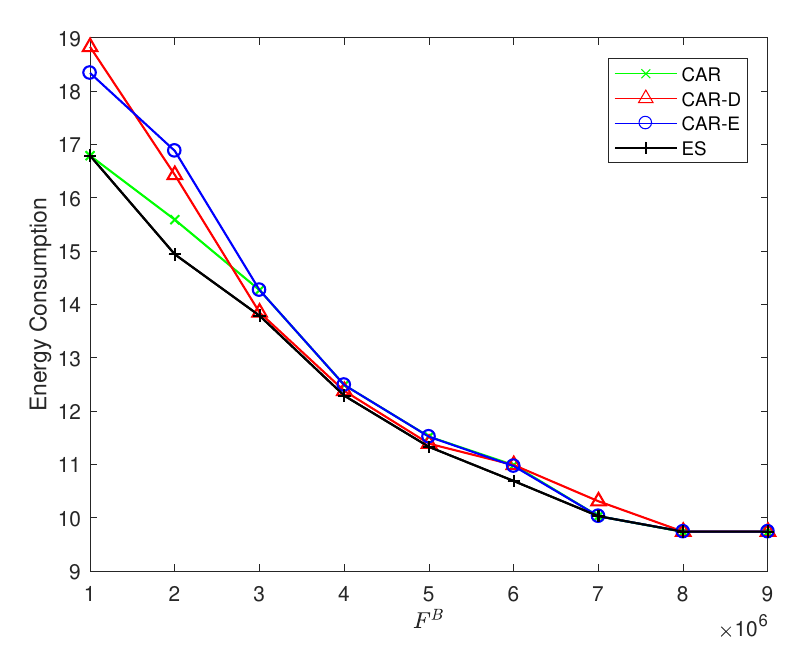}
	\caption{Sum energy consumption of all the UEs versus capacity of BBU pool, i.e., $F^B$.} \label{EEb}
\end{figure}

\begin{figure}[htbp]
	\centering
	\includegraphics[width=2.5in]{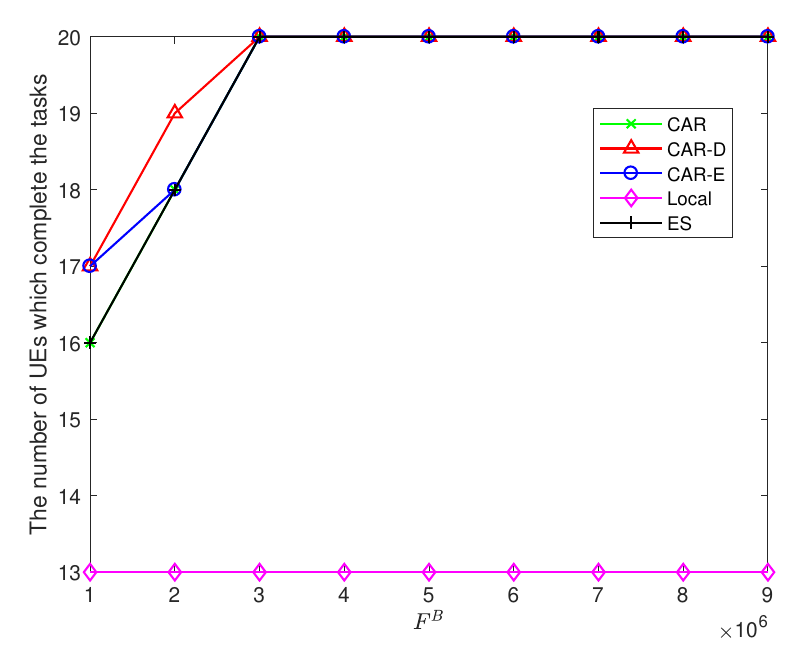}
	\caption{The number of UEs which can successfully complete the tasks versus capacity of BBU pool, i.e., $F^B$.} \label{OHb}
\end{figure}

\begin{figure}[htbp]
	\centering
	\includegraphics[width=2.5in]{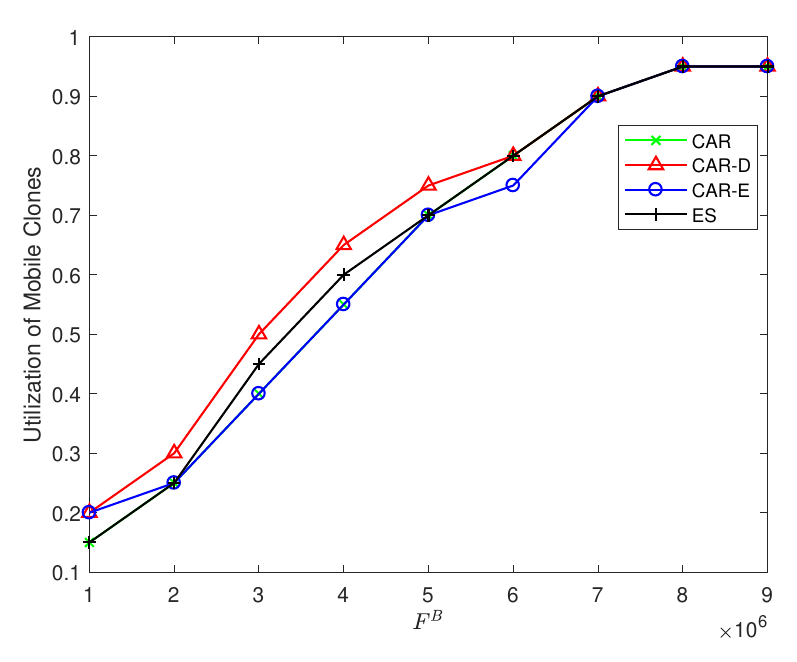}
	\caption{ Utilization of the mobile clones versus capacity of BBU pool, i.e., $F^B$.} \label{UCb}
\end{figure}

\begin{figure}[htbp]
	\centering
	\includegraphics[width=2.5in]{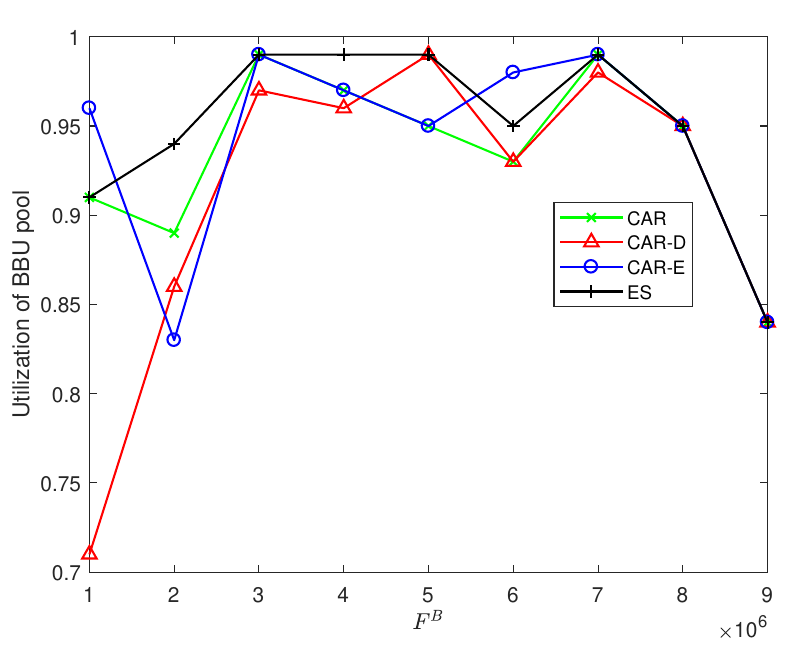}
	\caption{Utilization of the BBU pool versus capacity of BBU pool, i.e., $F^B$.} \label{UBb}
\end{figure}

\begin{figure}[htbp]
	\centering
	\includegraphics[width=2.5in]{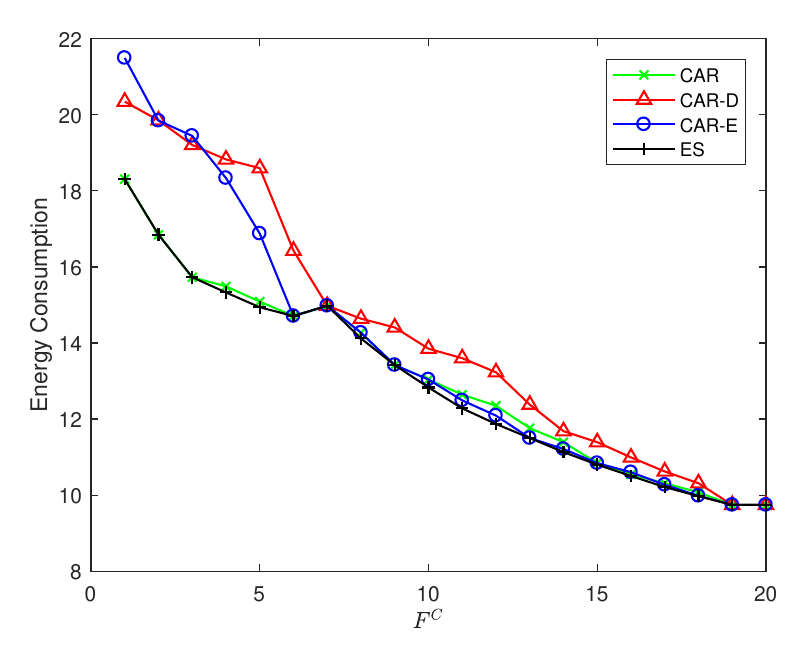}
	\caption{Sum energy consumption of all the UEs versus capacity of MC pool, i.e., $F^C$.} \label{EEc}
\end{figure}

\begin{figure}[htbp]
	\centering
	\includegraphics[width=2.5in]{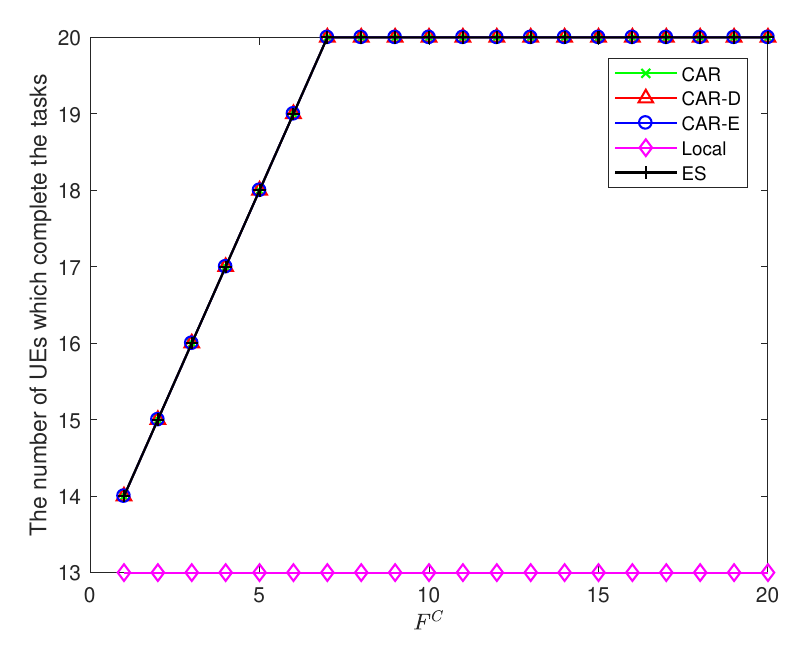}
	\caption{The number of UEs which can successfully complete the tasks versus capacity of MC pool, i.e., $F^C$.} \label{OHc}
\end{figure}

\begin{figure}[htbp]
	\centering
	\includegraphics[width=2.5in]{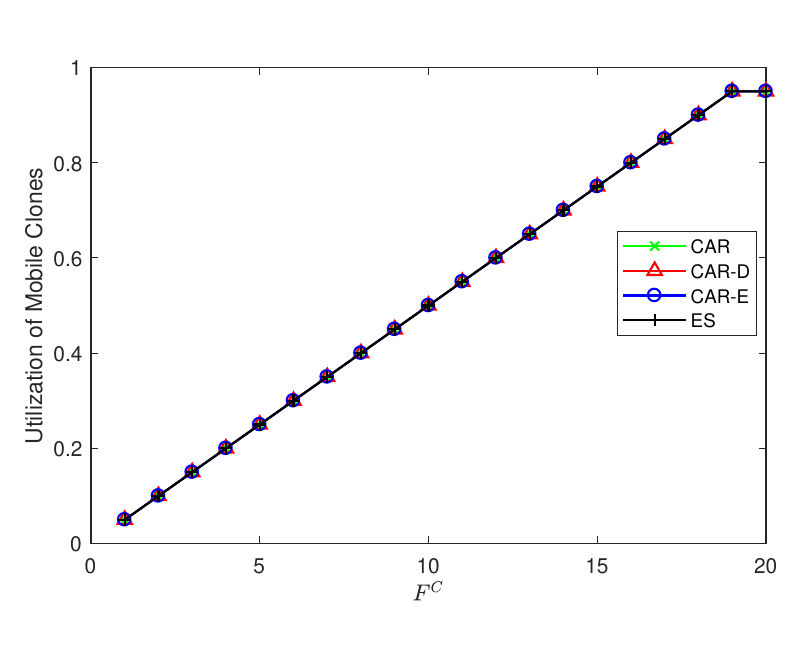}
	\caption{Utilization of the mobile clones versus capacity of MC pool, i.e., $F^C$.} \label{UCc}
\end{figure}

\begin{figure}[htbp]
	\centering
	\includegraphics[width=2.5in]{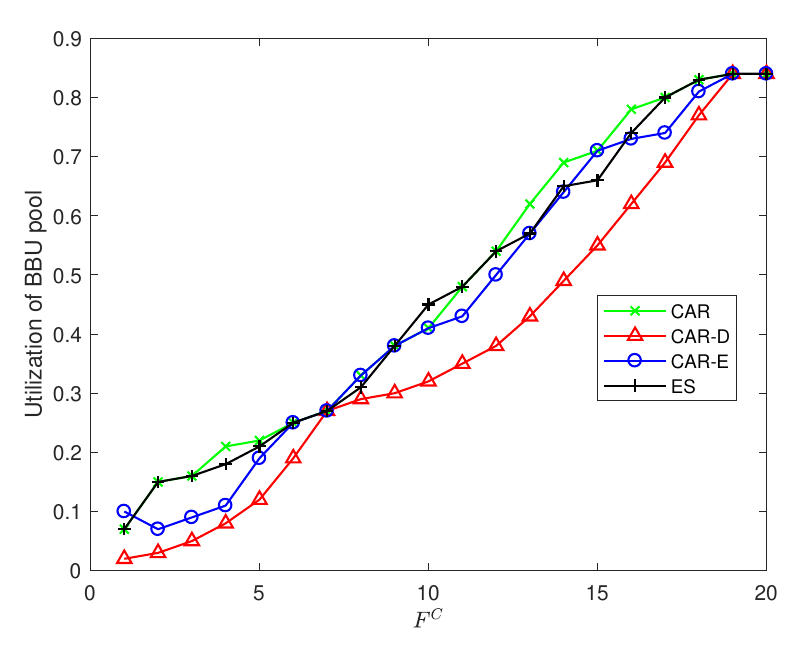}
	\caption{Utilization of the BBU pool versus capacity of MC pool, i.e., $F^C$.} \label{UBc}
\end{figure}

In Fig. \ref{EEb} - Fig. \ref{UBb}, $F^C=20$ is set, while in Fig. \ref{EEc} - Fig. \ref{UBc}, $F^B=9 \times 10^6$ cycles/s is assumed.

Fig. \ref{EEb} shows the sum energy consumption of all the UEs versus capacity of BBU pool, i.e., $F^B$. One can see that with the increase of the BBU capacity, sum energy consumption is decreased.
This is because as BBU capacity improves, more UEs can be allowed to offloading, which leads to more UEs benefiting from cloud and saving their energy. After the capacity of BBU pool reaches $8\times 10^6 $ cycles/s, the energy consumption keeps unchanged, as all the UEs required offloading are accepted. One can also see that ES achieves the smallest sum energy consumption, but as we mentioned before with prohibitive complexity. However, the performance of our proposed CAR, CAR-D and CAR-E are all every close to ES algorithm, especially with the increase of the BBU pool capacity. CAR can achieve the second best performance for the most of the examined values, with the help of 'while loop', as shown before. This operation will gradually remove the UE with the largest gap to its desired transmitting data rate from the offloading set.

Fig. \ref{OHb} shows the number of UEs which can successfully complete the tasks versus capacity of BBU pool.
One can see that with the increase of the BBU capacity, the number of UEs which can complete the tasks also increases. This is because with the increase of BBU capacity, more UEs which can not execute tasks before now can complete the computations in required time with the help of cloud. However, one sees that seven UEs can not complete the tasks via its own local execution.
One can also notice that after the BBU capacity reaches $3\times 10^6 $ cycles/s, no UEs fail in completing the tasks, either via local executing or offloading. When the BBU capacity is only $1\times 10^6 $ cycles/s, three UEs fail in completing tasks using CAR-D, whereas four UEs fail in finishing tasks using other algorithms. Surprisingly, our proposed low complexity algorithm, i.e.,  CAR-D make more UEs complete tasks
than other algorithms, including ES and our proposed CAR. This is because ES and CAR focusing on minimizing the sum energy consumption of all the UEs, thus resulting in some UE declined in offloading, where as CAR-D can accept more UEs, leading to a little bit higher sum energy consumption in some cases.

Fig. \ref{UCb} shows the utilization of the MC versus capacity of BBU pool, i.e., $F^B$, where the utilization of the MC is defined as that the number of accepted UEs over the whole capacity of the MC. One can see that with the increase of BBU capacity, the utilization increases as well until nearly one. This is because when BBU capacity improves, more UEs can be allowed to offload, leading to high utilization of MC. One sees that CAR-D achieves the best performance, even better than ES. This is because ES focuses on minimizing the sum energy consumption for all the UEs, thus may decline some UEs which contribute to increase of the whole energy consumption. However, CAR-D accepts the UEs with smallest required data first, until all the resource of BBU pool is used up, leading to high utilization of both BBU and MC pool.

Fig. \ref{UBb} shows utilization of the BBU pool versus capacity of BBU pool, where the utilization of the BBU pool is defined as all the processing data rate in BBU pool over the whole capacity of BBU pool. One can see that with the increase of BBU capacity, the utilization increases first until nearly one, and then drops. This is because when BBU capacity is small, most of the UEs are declined for offloading, resulting in low utilization of BBU pool. Also, when BBU capacity is large, the resource may be too much, leading to low utilization of BBU pool as well. As our objective is not to maximize the utilization, thus different algorithms may have different performance over different examined values. This figure can give us some insight on how to decide the BBU capacity, i.e., $F^B$ at the later stage. If one can proper decide the BBU capacity $F^B$, computation efficiency may be improved and resource waste can be avoid. However, to decide the proper $F^B$ is out of the scope of this paper.

Next, we will check the influence of the MC pool's resource to all the algorithms. Fig. \ref{EEc} shows the sum energy consumption of all the UEs versus capacity of MC pool, i.e., $F^C$. Similarly with before, one sees that with the increase of the recourse of MC pool, the sum energy consumption decreases for most of the cases, as expected.
One may note that the energy consumption at $F^C=7$ is a little bit higher than the case of $F^C=6$. This is because we assume that even for UEs in set $\mathcal{R}$, we still calculate their local energy consumption by using \textbf{Proposition 1} and assuming $f_i^{L^*}= f_{i,max}^L $, although their QoS requirement (i.e., constraint $C1$) is violated. However, it is easy to see that at the case of $F^C=7$, more UEs can meet their QoS requirement, although the energy consumption is a little bit high.
When the capacity of mobile clone reaches 19, the sum energy consumption keeps unchanged, as all the UEs requesting for offloading are accepted, therefore no more energy saving can be made.
Similarly with before, one sees that ES achieves the best performance and our proposed CAR is every close to ES, especially with the increase of the MC's capacity. CAR-D achieves the worst performance, as it focuses on adding the UE with the smallest data rate requirement into the offloading set first, leading to the largest offloading set. In another words, CAR-D normally makes more UEs benefiting from offloading, but not necessarily the minimum sum energy consumption for all the UEs. Surprisingly, CAR-E achieves smaller sum energy consumption for some examined values than CAR. This is because CAR-E accepts UE with largest saving energy first, leading to better performance than CAR in some cases. However, CAR can better balance the energy saving and the number of accepted UEs.

Fig. \ref{OHc} shows the number of UEs which can successfully complete the tasks versus capacity of MC pool. One can see that seven UEs fails in completing without the help of cloud (i.e., via Local execution). However, with the increase of the MC pool capacity, the number of UEs which can successfully complete the tasks increases, as expected. When the there is only one mobile clone (i.e., the MC capacity is 1), six UEs cannot complete the tasks, while when the capacity of MC pool reaches 7, no UEs failing in completing.

Fig. \ref{UCc} shows the utilization versus capacity of MC pool. One sees that with the increase of MC capacity, the utilization increases as well until nearly one, as expected. This is because when MC capacity increases, more UEs can be allowed to offload, leading to high utilization of MC pool.

Fig. \ref{UBc} shows utilization of the BBU pool versus capacity of MC pool. One can see that with the increase of MC capacity, the utilization increases until nearly one as well. CAR-D has the lowest utilization, for the reason that CAR-D accepts the UEs with the smallest required data rate first, resulting in low utilization.

%\begin{comment}
\section{Conclusion}
In this paper, we have proposed a novel ME-RAN architecture, which can support UEs' offloading and computation. Unified offloading framework has been presented and energy consumption minimization problem has been proposed to formulate as a non-convex mixed-integer optimization, which is hard to solve in general. The DLDA and CAR are introduced to deal with decision making and resource allocation, with the priority given to UEs which cannot complete the tasks locally. Two low complexity algorithms, i.e., CAR-E and CAR-D have also been proposed. Simulation results have been provided to show the effectiveness of the proposed architecture and algorithms. Future work will be focused on how to design the computation resource allocation algorithm between the BBU and MC in edge cloud.
%\end{comment}

\bibliographystyle{IEEEtran}
\bibliography{CAR}

\end{document}